\documentclass{mn2e} 
\usepackage{color}
\usepackage{dcolumn}
\usepackage{bm}
\usepackage{epsfig}

\def\reference{\par\noindent\hangindent\parindent}

 at 13 truept

\begin{document}

\twocolumn

\title[GAMA CSED]{Galaxy And Mass Assembly (GAMA): The $0.013<z<0.1$ cosmic spectral energy distribution from 0.1 $\mu$m to 1 mm \vspace{-0.7cm}}
\author[Driver et al.] 
{S.P.~Driver,$^{1,2}$\thanks{SUPA, Scottish Universities Physics Alliance}\thanks{e-mail:Simon.Driver@icrar.org},
A.S.G.~Robotham,$^{1,2}$
L. Kelvin,$^{1,2}$
M. Alpaslan,$^{1,2}$
I.K.~Baldry,$^3$
\newauthor 
S.P.~Bamford,$^4$
S.~Brough,$^5$
M.~Brown,$^{6}$,
A.M.~Hopkins,$^5$
J.~Liske,$^7$
J.~Loveday,$^8$
\newauthor 
P.~Norberg,$^9$
J.A.~Peacock,$^{10}$
E. Andrae,$^{11}$
J.Bland-Hawthorn,$^{12}$
N.~Bourne,$^4$
\newauthor
E.~Cameron,$^{13}$
M.~Colless,$^5$
C.J.~Conselice,$^4$
S.M.~Croom,$^{12}$
L.~Dunne,$^{14}$
\newauthor 
C.S.~Frenk,$^{9}$
Alister~W.~Graham,$^{15}$
M.~Gunawardhana,$^{12}$
D.T.~Hill,$^2$
D.H.~Jones,$^{6}$
\newauthor 
K.~Kuijken,$^{16}$
B. Madore,$^{17}$
R.C.~Nichol,$^{18}$
H.R.~Parkinson,$^{10}$
K.A.~Pimbblet,$^{6}$
\newauthor 
S.~Phillipps,$^{19}$
C.C.~Popescu,$^{20}$
M.~Prescott,$^3$
M. Seibert,$^{17}$,
R.G.~Sharp,$^{21}$
\newauthor 
W.J.~Sutherland,$^{22}$
E.N.~Taylor,$^{12}$
D.~Thomas,$^{18}$
R.J.~Tuffs,$^{11}$
E.~van~Kampen,$^7$
\newauthor
D.~Wijesinghe,$^{12}$
S. Wilkins$^{23}$\\
$^1$ICRAR\thanks{International Centre for Radio Astronomy Research}, The University of Western Australia, 35 Stirling Highway, Crawley, WA 6009, Australia\\
$^2$SUPA\thanks{Scottish Universities Physics Alliance}, School of Physics \& Astronomy, University of St Andrews, North Haugh, St Andrews, KY16 9SS, UK\\
$^3$Astrophysics Research Institute, Liverpool John Moores University, Egerton Wharf, Birkenhead, CH41 1LD, UK\\
$^4$Centre for Astronomy and Particle Theory, University of Nottingham, University Park, Nottingham NG7 2RD, UK\\
$^5$Australian Astronomical Observatory, PO Box 296, Epping, NSW 1710, Australia\\
$^6$School of Physics, Monash University, Clayton, Victoria 3800, Australia\\
$^7$European Southern Observatory, Karl-Schwarzschild-Str.~2, 85748 Garching, Germany\\
$^8$Astronomy Centre, University of Sussex, Falmer, Brighton BN1 9QH, UK\\
$^{9}$Institute for Computational Cosmology, Department of Physics, Durham University, South Road, Durham DH1 3LE, UK\\
$^{10}$SUPA, Institute for Astronomy, University of Edinburgh, Royal Observatory, Blackford Hill, Edinburgh EH9 3HJ, UK\\
$^{11}$Max Planck Institute for Nuclear Physics (MPIK), Saupfercheckweg 1, 69117 Heidelberg, Germany \\
$^{12}$Sydney Institute for Astronomy, School of Physics, University of Sydney, NSW 2006, Australia\\
$^{13}$ School of Mathematical Sciences, Queensland University of Technology, GPO Box 2434, Brisbane 4001, QLD, Australia \\
$^{14}$ Department of Physics and Astronomy, University of Canterbury, Private Bag 4800, Christchurch 8140, New Zealand \\
$^{15}$ Centre for Astrophysics and Supercomputing, Swinburne University of Technology, Hawthorn, Victoria 3122, Australia\\
$^{16}$Leiden University, P.O.~Box 9500, 2300 RA Leiden, The Netherlands\\
$^{17}$Observatories of the Carnegie Institution of Washington, 813 Santa Barbara Street, Pasadena, CA 91101, USA \\
$^{18}$Institute of Cosmology and Gravitation (ICG), University of Portsmouth, Dennis Sciama Building, Portsmouth PO1 3FX, UK\\
$^{19}$HH Wills Physics Laboratory, University of Bristol, Tyndall Avenue, Bristol, BS8 1TL, UK\\
$^{20}$Jeremiah Horrocks Institute, University of Central Lancashire, Preston PR1 2HE, UK\\
$^{21}$Research School of Astronomy \& Astrophysics, Mount Stromlo Observatory, Cotter Road, Western Creek, ACT 2611, Australia\\
$^{22}$Astronomy Unit, Queen Mary University London, Mile End Rd, London E1 4NS, UK \\
$^{23}$School of Physics and Astronomy, Oxford University, Keeble Road, Oxford, UK \vspace{-0.5cm}}

\maketitle

\label{firstpage}
\begin{abstract}
We use the GAMA I dataset combined with GALEX, SDSS and UKIDSS imaging
to construct the low-redshift ($z<0.1$) galaxy luminosity functions in
FUV, NUV, $ugriz$, and $YJHK$ bands from within a single well
constrained volume of $3.4 \times 10^5$ (Mpc/h)$^{3}$. The derived
luminosity distributions are normalised to the SDSS DR7 main survey to
reduce the estimated cosmic variance to the 5 per cent level. The data
are used to construct the cosmic spectral energy distribution (CSED)
from 0.1 to 2.1 $\mu$m free from any wavelength dependent cosmic
variance for both the elliptical and non-elliptical populations. The
two populations exhibit dramatically different CSEDs as expected for a
predominantly old and young population respectively. Using the Driver
et al.~(2008) prescription for the azimuthally averaged photon escape
fraction, the non-ellipticals are corrected for the impact of dust
attenuation and the combined CSED constructed. The final results show
that the Universe is currently generating $(1.8 \pm 0.3) \times
10^{35}$ h W Mpc$^{-3}$ of which $(1.2 \pm 0.1) \times 10^{35}$ h W
Mpc$^{-3}$ is directly released into the inter-galactic medium and
$(0.6 \pm 0.1) \times 10^{35}$ h W Mpc$^{-3}$ is reprocessed and
reradiated by dust in the far-IR.  Using the GAMA data and our dust
model we predict the mid and far-IR emission which agrees remarkably
well with available data. We therefore provide a robust description of
the pre- and post dust attenuated energy output of the nearby Universe
from 0.1$\mu$m to 0.6mm. The largest uncertainty in this measurement
lies in the mid and far-IR bands stemming from the dust attenuation
correction and its currently poorly constrained dependence on
environment, stellar mass, and morphology.
\end{abstract}

\setlength{\extrarowheight}{0pt}

\section{Introduction}
The cosmic spectral energy distribution (CSED) describes the energy
being generated within a representative volume of the Universe at some
specified epoch. See for example Hill et al.~(2010) for the most
recent empirical measurement, or Somerville et al.~(2012) for the most
recent attempt to model the CSED. Analogous to the baryon budget
(Fukugita, Hogan \& Peebles 1998), the CSED or energy budget provides
an empirical measurement of how the energy being produced in the
Universe at some epoch is distributed as a function of wavelength. The
CSED can be measured for a range of environments from voids to rich
clusters to follow the progress of energy production as a function
of local density. Furthermore, if one can measure the energy budget at
all epochs one effectively constructs a direct empirical blueprint of
the galaxy formation process, or at least its energy emission
signature (see for example Finke, Razzaque \& Dermer 2010). 

The CSED is a predictable quantity given a cosmic star-formation
history (CSFH; e.g., Hopkins \& Beacom 2006), an assumed initial mass
function (IMF; e.g., Kroupa 2002) along with any time (or other)
dependencies, and a stellar population model (e.g., {\sc pegase.2},
Fioc \& Rocca-Volmerange 1997, 2001). In practice knowledge of the
impact and evolution of dust and metalicity are also crucial and
discrepancies between the predicted and actual CSED can be used to
quantify these properties if the other quantities are considered
known. The CSED summed and redshifted over all epochs must also
reconcile with the sum of the resolved and unresolved extra-galactic
background (e.g., Gilmore et al.~2011 and references therein) modulo
some corrections for attenuation via the inter-galactic medium.  It
therefore represents a broadbrush consistency check as to whether many
of our key observations and assumptions are correct and whether
empirical datasets, often constructed in a relatively orthogonal
manner, are in agreement.

Traditionally the main focus for CSED measurements in the nearby
Universe is in the UV to far-IR wavelength range (Driver et
al.,~2008). This wavelength range is entirely dominated by starlight,
either directly (FUV to near-IR), or starlight reprocessed by warm
(mid-IR) or cold (far-IR) dust (see for example Popescu \& Tuffs 2002
or Popescu et al.~2011). At low redshift the contribution from other
sources (e.g., AGN) is believed to be negligible as is the
contribution to the energy budget from outside the FUV-FIR range (see
Driver et al.~2008). Note that this will not be the case at high
redshift where the incidence of AGN is much higher (e.g., Richards et
al.~2006), resulting in a possibly significant X-ray contribution to
the high-z CSED.  Note also that the Cosmic Microwave Background is
not considered part of the nearby CSED as although photons are passing
through the local volume, they do not originate from within it. Here
we take the CSED as specifically describing the instantaneous energy
production rate rather than the energy density which can be derived
from the CSED integrated over all time (redshifted, k-corrected, and
dust corrected appropriately).

The CSED is most readily constructed from the measurement of the
galaxy luminosity function from a large scale galaxy redshift survey
across a broad wavelength range. Measured luminosity functions in each
band provide an independent estimate of the luminosity density at one
specific wavelength, and when combined form the overall CSED. However,
one problem with this approach is that most surveys do not cover a
sufficiently broad wavelength range to construct the full
CSED. Instead the CSED has traditionally been constructed from a set
of inhomogeneous surveys which suffer from systematic offsets at the
survey wavelength boundaries. These offsets are difficult to quantify
and might be physical, i.e., sample/cosmic variance (Driver \&
Robotham~2011), or to do with the measurement process, e.g.,
incompleteness issues (Cross \& Driver~2002), or photometric
measurement discrepancies (e.g., Graham et al.~2005, see also Fig.~25
of Hill et al.~2011). For example SDSS Petrosian data might be
combined with 2MASS aperture photometry each with distinct biases in
terms of flux measurements. 

A discontinuity between the optical ($ugriz$) and near-IR data ($K$)
was first noted by Wright (2001) which was eventually traced to a
normalisation issue in the first SDSS luminosity functions. However an
apparent offset was also noted between the $z$ and $J$ bands by Baldry
\& Glazebrook~(2003) which remains unresolved. In Hill et al.~(2010)
we combined redshifts from the Millennium Galaxy Catalogue (Liske et
al.~2003) with imaging data from the Sloan Digital Sky Survey (York et
al.~2000) and the UK Infrared Deep Sky Survey Large Area Survey
(Lawrence et al.~2007) to produce a 9 band CSED ($ugrizYJHK$)
stretching from 0.3 to 2.1 $\mu$m. Although there was no obvious
optical/near-IR discontinuity the statistical errors were quite large
because of the small sample size. The six-degree field galaxy survey
(6dFGS; Jones et al.~2006) also sampled the optical and near-IR
regions ($b_Jr_FJHK$) within a single survey (Hill et al.~2010) and
again no obvious optical/near-IR discontinuity was seen (however the
6dF $r_F$ band data appears anomalously low compared to other $r$ band
measurements).

Here we use data from the Galaxy And Mass Assembly survey (GAMA;
Driver et al.~2009; 2011) to construct the CSED from FUV to near-IR
wavelengths within a single and spectroscopically complete volume
limited sample. The spectroscopic data were mainly obtained with the
Anglo-Australian Telescope (Driver et al.~2011), while the optical and
near-IR data is reprocessed SDSS and UKIDSS LAS imaging data, using
matched apertures. The photometry was performed on data smoothed to
the same resolution in each band (as described in Hill et
al.~2011). Hence while cosmic variance may remain in the overall
CSED amplitude, any wavelength dependence should be removed (modulo
any dependence on the galaxy clustering signature within the GAMA
volume).

In section 2 we describe the data and the construction of our
multi-wavelength volume limited sample. In section 3 we describe the
methodologies used to construct the luminosity functions and extract
the luminosity densities in 11 bands. In section 4 we apply the
methods to construct independent luminosity functions and the CSED
respectively. In section 5 we consider the issue of dust attenuation
which requires the isolation of the elliptical galaxies, believed to
be dust free (Rowlands et al.~2012), and the construction of the elliptical
(spheroid-dominated) and non-elliptical (disc-dominated)
CSEDs. Correcting the disc-dominated population using the photon
escape fractions given in Driver et al.~(2008) enables the
construction of the CSED both pre- and post attenuation. In section 6 we
derive the present energy output of the Universe. This is extrapolated
into the far-IR by calculating the attenuated energy and reallocating
this to an appropriate far-IR dust emission spectrum (e.g., Dale \&
Helou 2002) prior to comparison with available far-IR data. The
calibrated $z=0$ pre- and post- attenuated CSED from 0.1 to 1000
$\mu$m are available on request.

Throughout this paper we use $H_o$=100$h$km s$^{-1}$ Mpc$^{-1}$ and adopt
$\Omega_M=0.27$ and $\Omega_{\Lambda}=0.73$ (Komatsu et al.~2011).

\section{Data selection}
The GAMA I database (Driver et al.~2011) comprises 11-band photometry
from the GALEX satellite (FUV,NUV), reprocessed SDSS archival data
($ugriz$), reprocessed UKIDSS LAS archival data ($YJHK$), and
spectroscopic information (redshifts) from the AAT and other sources
to $r_{\rm pet}<19.4$ mag across three $4^\circ \times 12^\circ$
equatorial GAMA regions (see Driver et al.~2011 for a full description
of the GAMA survey; Baldry et al.~2010 for the spectroscopic target
selection, and Robotham et al.~2010 for the tiling
procedure). Photometry for the $ugrizYJHK$ bands is described in Hill
et al.~(2011) with revisions as given in Liske et al.~(in
prep.). Briefly all data are first convolved to a common $2''$ seeing.
Cutouts are made at the location of each galaxy in the GAMA input
catalogue. SExtractor is used to identify the central object and
measure its Kron magnitude. SExtractor in dual object mode is then
used to measure the flux in all other bands using the $r$-defined
aperture. Hence we achieve r-defined matched aperture photometry from
$u$ to $K$. A complete description of the GAMA $u$ to $K$ photometry
pipeline is provided in Hill et al.~(2011) and has been shown to
produce improved colour measurements over archival data in all
bands. Here we use version 2 photometry (see Liske et al.~in prep.) in
which the uniformity of the convolved PSF across the $ugrizYJHK$ data
was improved, and some previously poor quality near-IR frames
rejected.

Star-galaxy separation was implemented prior to the spectroscopic
survey, as defined in Baldry et al.~(2010), and used a combination of
size and colour cuts. The additional optical-near-IR colour selection
process was demonstrated to be highly effective in recovering compact
galaxies with fairly minimal stellar contamination. The GALEX data is
from a combination of MIS archival and proprietary data obtained by
the MIS and GAMA teams. The GALEX photometry uses independent software
optimised for galaxy source detection and flux measurement, and is
described in detail in Seibert et al.~(in prep.). The GALEX data is
matched to the $r$-band defined catalogue following the method
described in Robotham \& Driver (2011) whereby flux is redistributed
for multiple-matched objects according to the inverse of the first
moment of the centroid offsets. Matches are either: unambiguous
(single match, 46 per cent; no UV detection, 31 per cent), or
redistributed between two (16 per cent), three (5 per cent), four (1
per cent), or more (1 per cent) objects.  

The spectroscopic survey to $r_{\rm Pet} < 19.4$ mag --- which are
predominantly acquired using the AAOmgea prime-focus fibre-fed
spectrographs on the Anglo-Australian Telescope --- are complete to
$97.0$ per cent with no obvious spatial or other bias (see Driver et
al.~2011). Redshifts for the spectra are assigned manually and a
quality flag allocated. Following a calibration process to a standard
quality system only $nQ \geq 3$ redshifts are used which implies a
probability of being correct of $>0.9$ (see Driver et al.~2011). The
redshift accuracy from repeat observations is known to be $\sigma_{\rm
  v}=\pm 65$ km/s declining to $\sigma_{\rm v}=\pm 97$ km/s for the
lowest signal-to-noise data (see Liske et al.~in prep.).

~

\noindent
The exact internal GAMA I catalogues extracted from the database and
used for this paper are:

~

\noindent
{\sf TilingCatv16} --- cataid\footnote{cataid is the unique GAMA I identifier}, Right Ascension, Declination, and redshift quality (see Baldry et al.~2010)

\noindent
{\sf DistanceFramesv06} --- flow corrected redshifts (see Baldry et al.~2012.)

\noindent
{\sf ApMatchedPhotomv02} --- $ugrizYJHK$ Kron aperture matched photometry (see Hill et al.~2011; Liske et al.~in prep.)

\noindent
{\sf GalexAdvancedmatchV02} --- FUV and NUV fluxes positional matching with flux redistribution (see Seibert et al.~in prep.)

\noindent
{\sf SersicCatv07} --- $r$-band Sersic indices (see Kelvin et al.~2012.)

\begin{figure}
\centerline{\psfig{file=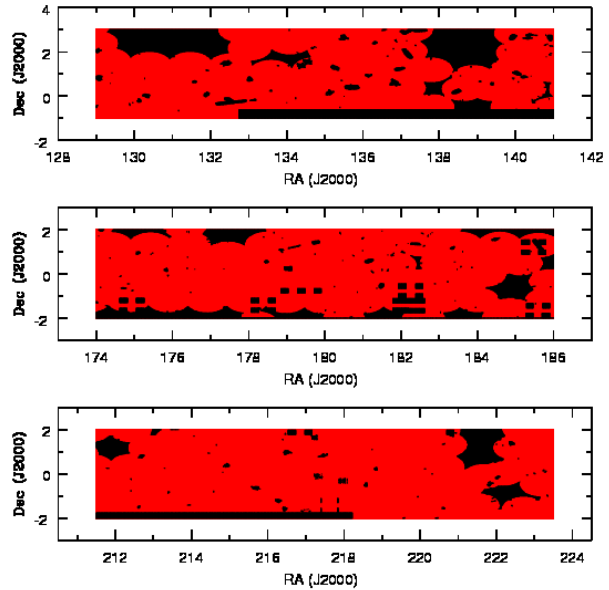,width=\columnwidth}}
\caption{\label{fig:imagemap} The area of G09 (top), G12 (centre) and
  G15 (bottom) surveyed in all 11 filters. The full GAMA I regions are
  shown in black and the common region subset in red. The total
  coverage is $125.06$deg$^2$, see Table~\ref{tab:coverage} for the
  coverage in each band.}
\end{figure}

\subsection{Extracting a common region}
At the present time imaging coverage of the GAMA regions in all
11-bands is incomplete. In addition there are a number of exclusion
regions where galaxies could not be detected due to bright stars
and/or defects in the original SDSS imaging data. However, by far the
main reason for the gaps are individual UKIDSS LAS pointings failing
quality control and/or the need for GALEX to avoid very bright
stars. In order to derive the 11-band CSED we must identify an area
over which complete photometry can be obtained, and the appropriate
sky coverage of this region.  Using the formula given in Driver \&
Robotham (2010; Eqn.~4) we find that the $1\sigma$ sample/cosmic
variance in three individual GAMA pointings with $z<0.1$ to be 14 per
cent. We estimate the coverage of the complete 11-band common region
by sampling our SWARPed mosaics at regular 1 arcminute intervals over
the three GAMA I regions and measuring the background value at each
location. A value of zero (in the case of the $ugrizYJHK$ SWARPs) or a
value of less than $-10$ (for the FUV and NUV data) indicates no data
at the specified location. We then combine the 11 independent coverage
maps to obtain the combined coverage map as shown in
Fig.~\ref{fig:imagemap}, and highlighting the complexity of the
mask. Note that for clarity we do not include the bright star mask or
SDSS exclusion mask which diminishes the area covered by a further 0.9
per cent in all bands.

The implied final survey area, which includes the common region minus
masked areas, for our combined $FUVNUVugrizYJHK$ catalogue is
therefore $125.06$deg$^2$. providing a catalogue containing 80464
objects to a uniform limit of $r_{\rm Kron} < 19.4$ mag of which
redshifts are known for 97.0 per cent (see Table~\ref{tab:limits} for
completeness in other bands).

\begin{table}
\caption{Coverage of the GAMA regions by filter, for a common area in
  all filters (All), within each GAMA sub-region (G09, G12, G15), or
  for the three regions combined (GAMA). Errors throughout are
  estimated to be $< \pm 1$ per cent. \label{tab:coverage}}
\begin{center}
\begin{tabular}{c|r|r|r|r} \hline
Filter & G09 & G12 & G15 & GAMA \\ 
       & (\%) & (\%) & (\%) \\ \hline
FUV& 83.1& 89.5& 93.9& 88.8 \\
NUV& 84.3& 90.0& 93.9& 89.4 \\
u  &100.0  &100.0  &100.0  &100.0   \\
g  &100.0  &100.0  &100.0  &100.0   \\
r  &100.0  &100.0  &100.0  &100.0   \\
i  &100.0  &100.0  &100.0  &100.0   \\
z  &100.0  &100.0  &100.0  &100.0   \\
Y  & 93.1& 96.2& 96.1& 95.2 \\
J  & 93.1& 96.2& 96.1& 95.2 \\
H  & 96.8& 96.2& 99.6& 97.5 \\
K  & 96.8& 96.2& 99.5& 97.5 \\ 
All& 76.8& 86.5& 90.1& 84.5 \\ \hline
\end{tabular}
\end{center}
\end{table}

\subsection{Selection limits in each band}
GAMA version 2 matched aperture photometry is derived only for
galaxies listed in the GAMA I input catalogue which includes multiple
flux selections (see Baldry et al.~2010). This catalogue is then
trimmed to a uniform spectroscopic survey limit of $r_{\rm Kron} <
19.4$ mag. This abrupt $r$-band cut naturally introduces a colour bias
in all other bands, making the selection limits in each band dependent
on the colour distribution of the galaxy population.

\begin{figure*}

\centerline{\psfig{file=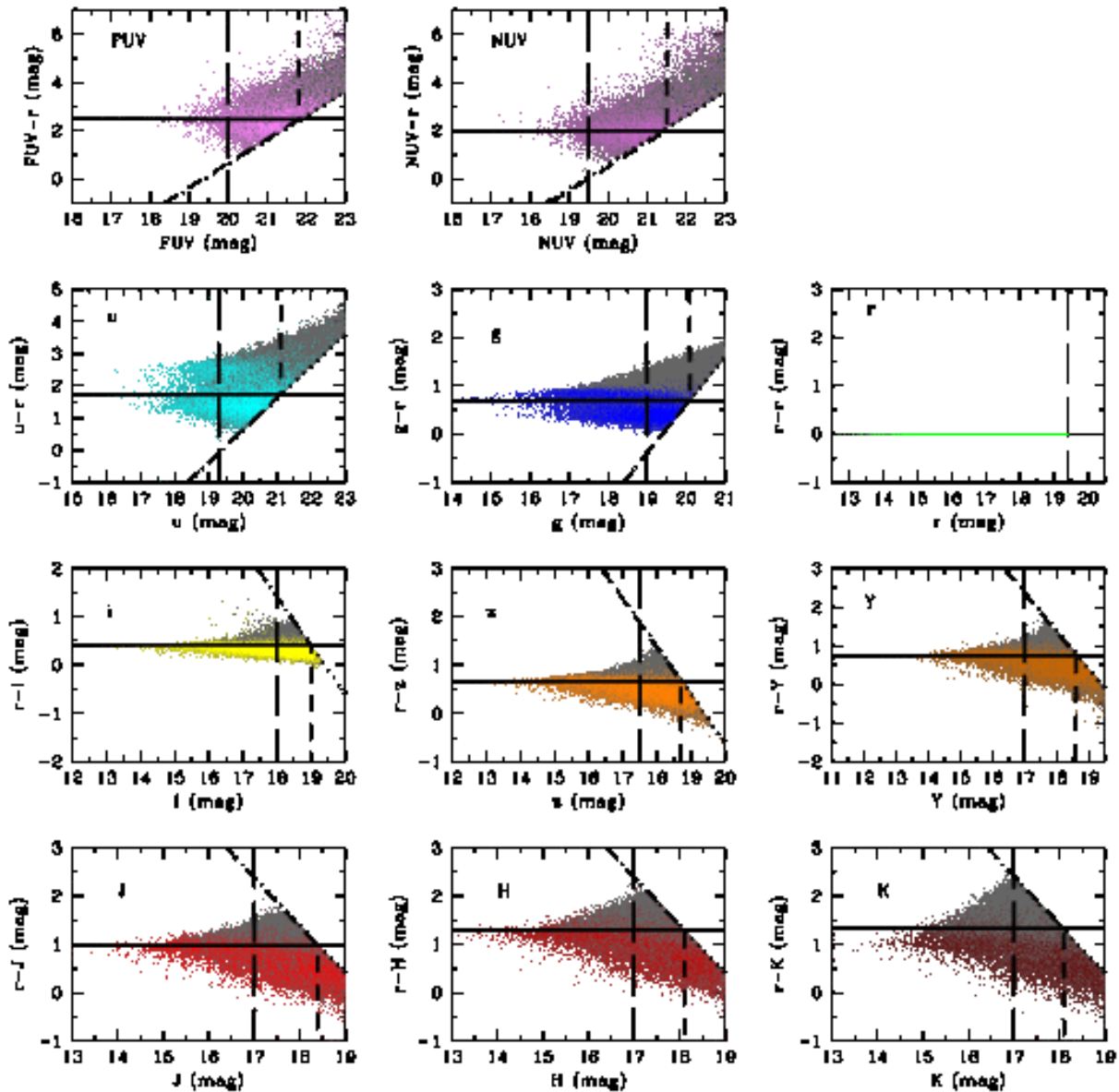,width=\textwidth}}

\caption{\label{fig:mcol} The colour distributions for our 11 band
  data w.r.t $r$-band. The various lines show the three selection
  boundaries discussed in the text with the thick medium dashed
  boundary taken as our final $1/V_{\rm Max}$ limit. Data in the redshift range
  $0.013 < z < 0.1$ are shown in colour and these are the data used in
  this paper to derive the CSED.}
\end{figure*}

\begin{figure*}

\vspace{-2.0cm}

\centerline{\psfig{file=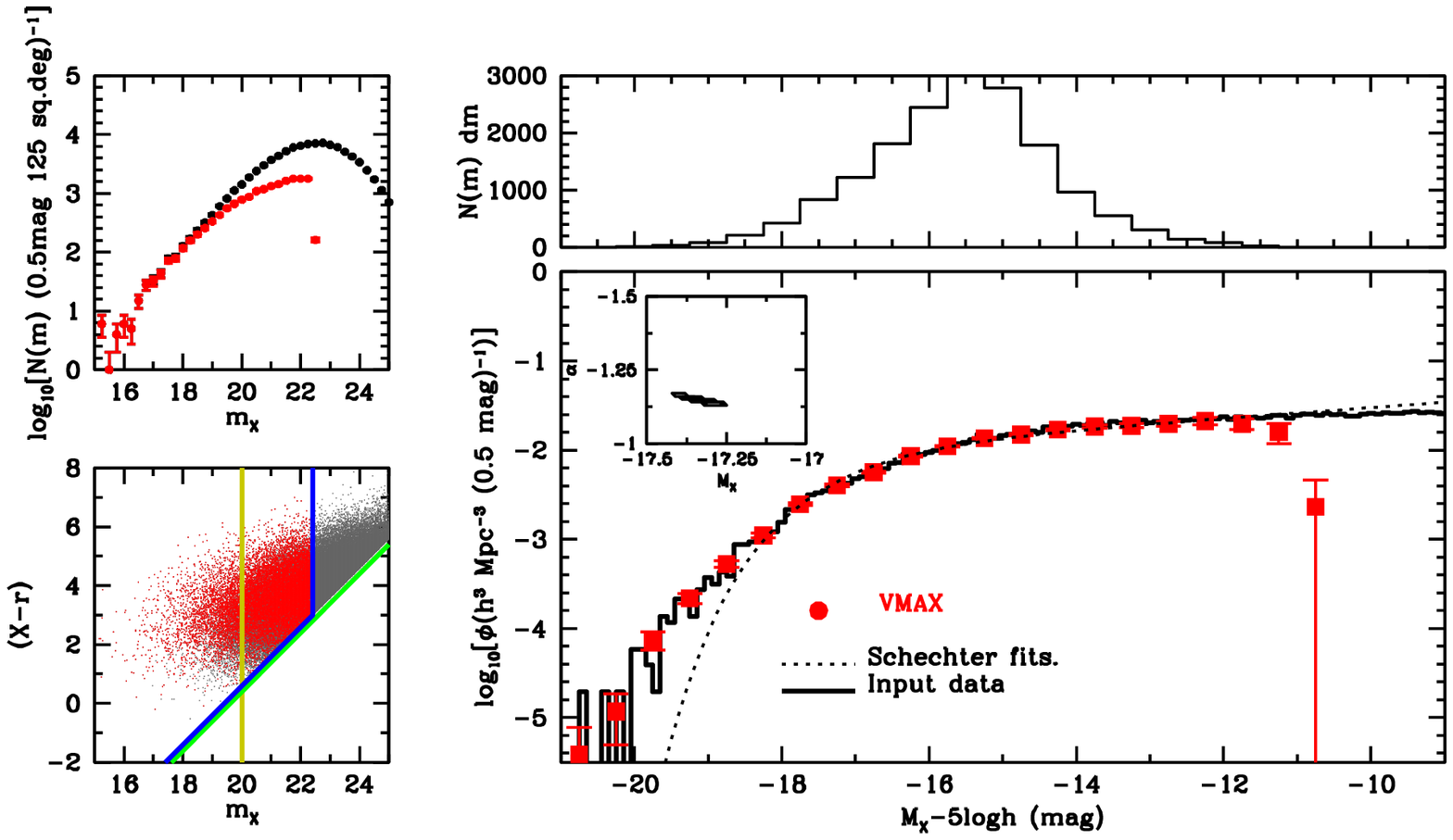,width=\textwidth}}

\vspace{-4.0cm}

\caption{\label{fig:testlf} ({\it main panel}) An illustration of the
  accuracy of our luminosity density estimator ($1/V_{\rm Max}$, red
  squares) as compared to the input test data (solid black
  histogram). Also shown is the standard Schechter function fit. ({\it
    main panel inset}) The 1-,2- and 3- $\sigma$ error contours for
  the best fit Schechter function to the $1/V_{\rm Max}$ data.  ({\it
    upper panel}) The actual number of galaxies used in the derivation
  of the luminosity distributions. ({\it upper left}) The galaxy
  number-counts prior to any flux or redshifts cuts (black data
  points) and after the flux and redshift cuts as indicated in
  Table~\ref{tab:limits}. ({\it lower left}) The colour-magnitude
  diagram showing the full data set prior to cuts (black dots) and
  after flux and redshift cuts (coloured dots). The coloured lines
  denote the various selection boundaries as described in section~2.2.}
\end{figure*}

To identify appropriate selection limits we show on
Fig.~\ref{fig:testlf} the colour magnitude diagrams for our data in
each band.  Following Hill et al.~(2010) we identify three obvious
selection boundaries for this dataset: (1) The limit at which a colour
unbiased catalogue can be extracted (long dashed lines). (2) a colour
dependent limit which traces the colour-bias (dotted lines). (3) a
colour dependent limit until the mean colour is reached after which a
constant limit is enforced (short dashed lines). Volume-corrected
luminosity distributions can be determined within each of these limits
with varying pros and cons. For example while limit 1 offers the simplest
and most secure route it uses the minimum amount of data increasing
the random errors and susceptibility to cosmic variance. Limit 2 uses
all the data but much of these data lie at very faint flux limits
which are prone to large photometric error, and the shape of the
boundary renders the results particularly susceptible to Eddington
bias. Limit 3 represents a compromise between utilising excessive poor
quality data and reducing the sample excessively. This limit was
adopted in Hill et al.~(2010) and we follow this practice here.  The
relevant limits, resulting sample sizes, and spectroscopic
completeness in each band are shown on Table~\ref{tab:limits}. On
Fig.~\ref{fig:mcol} data in the redshift range $0.013 < z < 0.1$ are
shown as coloured dots.

\begin{table*}
\caption{Data selection process. Column 2 shows the limit above which the
  sample is entirely free of any colour bias. Column 3 shows the mean
  colour above this limit and column 4 shows the derived faint limit
  which we adopt in our LF analysis and is defined as the flux at
  which the sample becomes incomplete for the mean colour. The
  remaining columns show the sample size, spectroscopic completeness,
  and the final number of galaxies used in the luminosity function
  calculation once all selection limits are
  imposed. \label{tab:limits}}
\begin{center}
\begin{tabular}{c|c|c|c|c|c|c} \hline
Filter & Bright limit & Mean colour & Faint limit & No & Comp. & No ($0.013<z<0.1$)\\
     & (AB mag)     &  $(X-r)$    & (AB mag)    &            & (\%) \\ \hline 
 FUV & 20.0 & $2.46 \pm 0.86$ & 21.8 & 21740 & 98.2 & 7210 \\
 NUV & 19.5 & $2.01 \pm 0.90$ & 21.5 & 30247 & 97.9 & 7989 \\
 $u$ & 19.3 & $1.73 \pm 0.58$ & 21.1 & 48602 & 97.8 & 10463 \\
 $g$ & 19.0 & $0.71 \pm 0.27$ & 20.1 & 60893 & 97.7 & 10990 \\
 $r$ & 19.4 & N/A & 19.4 & 80464 & 97.0 & 11032 \\
 $i$ & 18.0 & $-0.42 \pm 0.09$ & 19.0 & 77586 & 97.1 & 10609 \\
 $z$ & 17.5 & $-0.66 \pm 0.16$ & 18.7 & 72821 & 97.3 & 9756 \\
 $Y$ & 17.3 & $-0.75 \pm 0.21$ & 18.6 & 68156 & 97.5 & 9078 \\
 $J$ & 17.3 & $-0.98 \pm 0.26$ & 18.4 & 66249 & 97.5 & 8340 \\
 $H$ & 17.0 & $-1.29 \pm 0.29$ & 18.1 & 66428 & 97.5 & 8172 \\
 $K$ & 16.8 & $-1.33 \pm 0.39$ & 18.1 & 67227 & 97.5 & 7638 \\ \hline
\end{tabular}
\end{center}
\end{table*}

\section{Method}

\subsection{Luminosity distribution estimation}
In order to derive volume-corrected luminosity distributions in each
band we adopt a standard $1/V_{\rm Max}$ method, this is preferred
over a step-wise maximum likelihood method as it can better
accommodate the use of multiple selection limits to overcome colour
bias. The standard $1/V_{\rm Max}$ method (Schmidt 1968) can be used
to calculate the volume available to each galaxy based on its $r$ and
$X$ magnitude limits, i.e., $X_{\rm Lim}$ is the brightest of
$19.4-(r-X)$ or $X_{\rm Faint Limit}$ (where $X$ represents FUV, NUV,
$ugizYJH$ or $K$), and the selected redshift range ($0.013-0.1$). It
is worth noting that because of the depth of the survey and the
restricted redshift range that this generally constitutes a
volume-limited sample at the bright-end of the recovered luminosity
distribution in all bands, typically extending $\sim 2$ mags below
$L^*$. Using the $1/V_{\rm Max}$ estimator the luminosity distribution
is given by:

~

\noindent
$\phi(M) = \frac{C_X}{\eta}.\sum_{i=1}^{i=N} (\frac{1}{V_i})$

~

\noindent
where the sum is over all galaxies with $M-0.25 < M_i < M+0.25$,
$\eta$ is the cosmic variance correction for the combined GAMA data
over this redshift range (taken here as 0.85, see Driver et al.~2011,
Fig.~20), and $C_X$ is the inverse incompleteness given by
$C_X=\frac{N({\rm all})}{N({\rm with redshifts})}$. Note that the
incompleteness is handled in this simplistic way because it is so low
$<3.0$ per cent in all bands (see Table~\ref{tab:limits}).

Our $1/V_{\rm Max}$ estimator has been tested on trial data and the
results are shown on the main panel of Fig.~\ref{fig:testlf}. These
data were constructed using an input $r$-band Schechter function of
$M_r^* - 5\log_{10}h=-20.5$, $\alpha=-1.00$ and $\phi_*=0.003$
(Mpc/h)$^{-3}$, and used to populate a $125$deg$^2$ volume to
$z=0.1$. Colours were allocated assuming a Gaussian colour
distribution with offset $(X-r)=3.00$ and $\sigma=1.00$.  The test
sample was then truncated to $r_{\rm Kron} < 19.4$ mag and the method
described above used to recover the luminosity distribution (as
indicated by the solid red data points on Fig.~\ref{fig:testlf}). The
figure indicates that the luminosity distribution recovered is
accurate (within errors) as long as $>10$ galaxies are detected within
a magnitude bin. It is worth highlighting that although the test data
were drawn from a perfect Schechter function distribution, the
transformation to a second bandpass under the assumption of a Gaussian
colour distribution causes the distribution to become non-Schechter
like. As a consequence the bright-end of our test data, when shown in
the transformed bandpass, is poorly fitted by a Schechter function (as
indicated by the dotted line) with implications for the derived
luminosity density as discussed in the next section. The rather
obvious conclusion is that one should not expect a Schechter function
to fit in all bandpasses as the colour distribution between bands
essentially acts as a broad smoothing filter. This is compounded by
variations in the colour distribution with luminosity/mass and type as
well as the impact of dust attenuation which will likewise smear the
underlying distributions in the bluer bands (see Driver et al.~2007).

Finally we note that the method described above manages the colour
bias by increasing or reducing the $1/V_{\rm Max}$ weighting according
to each objects colour. This will ultimately break-down at the low
luminosity end when a galaxy of a specific colour becomes entirely
undetectable at our lower redshift limit of $z \sim 0.013$. We can
estimate this limit by asking what is the absolute magnitude of a
galaxy with the bright limit indicated in Table~\ref{tab:limits}
located at $z=0.013$. These values are: -13, -13.5, -13.5, -14, -14,
-15, -15.5, -15.8, -15.8, -16, and -16.2 for $FUV,NUV,ugrizYJHK$
respectively and we adopt these faint absolute magnitude limits when
fitting for the Schechter function parameters.

\subsection{Luminosity density ($j_{\lambda}$)}
In this paper we derive two measurements of the luminosity density.
The first is from the integration of the fitted Schechter function and
the second is from a direct summation of the $1/V_{\rm Max}$ weighted
fluxes. Both methods have their merits and weaknesses.

~

\noindent
{\bf Method 1: Schechter function fitting:} The $1/V_{\rm Max}$ data is
fitted by a simple three-parameter Schechter function (Schechter 1976)
via standard $\chi^2$-minimisation. The luminosity density is then
derived from the Schechter function parameters in the usual way
($j_X=\phi^*_XL^*_X\Gamma(\alpha_X+2)$ where $X$ denotes filter). This
is perhaps the most standard way of calculating the luminosity density,
but it extrapolates flux to infinitely large
and small luminosities. In particular, galaxy luminosity functions
often show an upturn at both bright and faint luminosities and unless
more complex forms are adopted the faint-end in particular is rarely a
good fit (see for example the unrestricted GAMA $ugriz$ luminosity
functions with a focus on the faint-end slopes reported in Loveday et
al.~2012). Non-Schechter like form is often seen at the very
bright-end as well, particularly in the UV and NIR wavebands (see for
example Robotham \& Driver~2011 or Jones et al.~2006). The errors for
Method 1 are derived by mapping out the full 1-sigma error ellipse in
the $M^*$-$\alpha$ plane having already optimised the normalisation at
each location within this ellipse. The error is then the largest
offset in $M^*$ or $\alpha$ within this $1\sigma$ error ellipse.

~

\noindent
{\bf Method 2: $1/V_{\rm Max}$ Summation:} One can directly sum the
$1/V_{\rm Max}$ weighted luminosities from the individual galaxies
within the selection boundaries i.e., ($j_X=\sum^{i=N}_{i=0}
10^{-0.4(M_i-M_{\odot})}/V_i$). This does not include any
extrapolation but rather assumes that the galaxy luminosity
distribution is fully sampled over the flux range which contributes
most to the luminosity density. The errors are derived from the
uncertainty in the flux measurements which we take from Hill et
al.~(2010) to be $\pm 0.03$ mag in the $griz$, $\pm 0.05$ in the
$uYJHK$ bands and $\pm 0.1$ mag in the NUV and FUV (from the mag error
distribution given in the {\sc GalexAdvancedmatchv02} catalogue).

~

\noindent
For our test data the known value is $3.0 \times
10^9$L$_{\odot}$Mpc$^{-3}$ and both methods recover accurate
measurements of the underlying luminosity density:

~

Method 1: $(3.03^{+0.20}_{-0.24}) \times 10^9$L$_{\odot}$Mpc$^{-3}$,

Method 2: $(2.97^{+0.15}_{-0.18}) \times 10^9$L$_{\odot}$Mpc$^{-3}$,

~

\noindent
This is perhaps surprising given the apparently poor fit of the
Schechter function to the bright end of the data
(Fig.~\ref{fig:testlf}; main panel) and indicates how strongly the
integrated luminosity density depends on the $L^*$ population. For
this paper we will adopt Method 2 as our preferred luminosity density
measurements for two main reasons: (1) it includes no extrapolation,
and (2) it most closely mirrors the actual distribution over the
region which dominates the luminosity density ($M^* \pm 2$ mag).

\subsection{Cosmic energy density ($\epsilon$)}
Our luminosity densities ($j_{\lambda}$) are by convention quoted in
units of $L_{\odot, \lambda} h$ Mpc$^{-3}$. To convert to an energy
density which represents the instantaneous energy production rate we
need to multiply by the effective mean frequency of the filter in
question (as given by the pivot wavelength, $\lambda_p$). We then
convert from solar units ($L_{\odot}$) to luminosity units ($W
Hz^{-1}$). This is achieved using the formula below, where the
observed energy density, $\epsilon^{\rm Obs}$, is given in units of W
$h$ Mpc$^{-3}$:
\begin{equation}
\epsilon^{\rm Obs} = \frac{c}{\lambda}j_{\lambda}10^{-0.4(M_{\odot,\lambda}-34.10)}
\end{equation}
the constant term of 34.10 is that required to convert AB magnitudes
to luminosity units (i.e., following the Oke \& Gunn 1983 definition
of the AB magnitude scale in which $m_{AB}=-2.5\log_{10}f_{\nu}+56.1$,
i.e., $m_{AB}=0$ when $f_{\nu}=3.631 \times 10^{-23}$W m$^{-2}$
Hz$^{-1}$, and $F_{\nu}=4 \pi d^2 f_{\nu}$ where $d$ is the standard
calibration distance of 10pc). The observed energy density,
$\epsilon^{\rm Obs}$, can be converted to an intrinsic energy density,
$\epsilon^{\rm Int}$, using the mean photon escape fraction ($p_{\rm
  esc, \lambda}$) defined in Driver et al.~(2008, Fig.~3), i.e.,
\begin{equation}
\epsilon^{\rm Int} = \epsilon^{\rm Obs}/p_{{\rm esc},\lambda}.
\end{equation}
the values adopted for the fixed parameters and their associated
errors are shown in Table~\ref{tab:params}. Note that although the
solar absolute magnitude is required in Eqn.~1 this is only because of
the convention of reporting the luminosity density, $j_{\lambda}$, in
units of $L_{\odot, \lambda}$ Mpc$^{-3}$ as an intermediary step. We
adopt this practice to allow for comparisons to previous work but note
that the final energy densities are not dependent on the solar
luminosity values used. More formally we define the luminosity density
to be:
\begin{equation}
j_{\lambda}=\phi_* 10^{-0.4(M^*_{\lambda}-M_{\odot, \lambda})}\Gamma(\alpha+2)
\end{equation}
for method 1, where $\phi^*, M^*$ and $\alpha$ are the usual Schechter function parameters, and
\begin{equation}
j_{\lambda}=\displaystyle\sum\limits^{n=i}_{n=1} (10^{-0.4(M_{i, \lambda}-M_{\odot})}/V_{\rm Max, i})
\end{equation}
for method 2, where $M_i$ represents the absolute magnitude of the
$i^{th}$ object within the specified flux limits and $V_{\rm Max, i}$
is the maximum volume over which this galaxy could have been seen.

\begin{table}
\caption{Various constants required for calculation of the luminosity
  density and energy densities. \label{tab:params}}
\begin{center}
\begin{tabular}{c|l|r|r|r} \hline
Filter & $A_\lambda /A_r$$^{\dagger}$ & $\lambda_{\rm Pivot}$ & $M_{\odot}^{\ddagger}$ & $p_{\rm esc}$ \\
       &  & (\AA) & (AB mag) & (\%) \\ \hline
FUV    & 3.045 &  1535 & 16.02 & $23 \pm 6$ \\
NUV    & 3.177 &  2301 & 10.18 & $34 \pm 6$ \\
u      & 1.874 &  3557 &  6.38 & $46 \pm 6$ \\
g      & 1.379 & 4702 &  5.15 & $58 \pm 6$ \\
r      & 1     & 6175 &  4.71 & $59 \pm 6$ \\
i      & 0.758 & 7491 &  4.56 & $65 \pm 6$ \\
z      & 0.538 & 8946 &  4.54 & $69 \pm 5$\\
Y      & 0.440 & 10305 &  4.52 & $72 \pm 5$\\ 
J      & 0.323 & 12354 &  4.57 & $77 \pm 4$ \\
H      & 0.210 & 16458 &  4.71 & $82 \pm 4$\\
K      & 0.131 & 21603 &  5.19 & $87 \pm 3$ \\ \hline
\end{tabular} 
\end{center}

$^{\dagger}$ values taken from Liske et al.~(in prep.).

$^{\ddagger}$ values taken from Hill et al.~(2011) for $u$ to $K$ and for FUV and NUV from http://www.ucolick.org/$\sim$cnaw/sun.html.
\end{table}

\section{Derivation of the luminosity distributions and densities}

\subsection{Corrections to the data}
Before the methodology described in the previous section can be
implemented a number of corrections to the data must be made to
compensate for practical issues of the observing process and known
systematic effects.

\subsubsection{Galactic extinction and flow corrections}
The individual flux measurements in each band are Galactic extinction
corrected using the Schlegel maps (Schlegel, Finkbeiner \& Davis 1998)
with the adopted $A_v$ terms for the 11 bands listed in
Table~\ref{tab:params}. The individual redshifts are also corrected
for the local flow as described in Baldry et al.~(2012). These taper
the Tonry et al.~(2000) multi-attractor model adopted at very low
redshift ($z<0.02$) to the CMB rest frame out to $z=0.03$ (see also
Loveday et al.~2012). This correction has no significant impact on the
results in this paper but see Baldry et al.~(2012) for discussions on
the effect this has on the low mass end of the stellar mass function.

\subsubsection{Redshift incompleteness}
The redshift incompleteness for each sample is shown in
Table~\ref{tab:limits}. As the spectroscopic completeness is
exceptionally high ($>97$ per cent in all filters), it is not
necessary to model the trend with magnitude, and so all results are
simply scaled up by the incompleteness values. The caveat is that a
highly biased incompleteness could have an impact at the very
faint-end where the volumes sampled are exceptionally small. However,
as we shall see the luminosity density is entirely dominated by $L^*$
systems and small variations in the derived luminosity density at the
very faint-end will have a negligible impact on the CSED.

\subsubsection{Absolute normalisation and sample/cosmic variance}
In Driver et al.~(2011) it was reported that the combined GAMA
coverage to $z<0.1$ is 15 per cent underdense with respect to the SDSS
main survey. This was estimated by comparing the number of r-band
$L^*$ galaxies in the GAMA volume to that in the SDSS Main Survey NGP
region. We therefore bootstrap to the larger SDSS area by rescaling
all normalisation values upwards by 15 per cent to accommodate for
this underdensity. We note that by recalibrating the $L^*$ density to
the SDSS main survey we reduce the cosmic variance in the GAMA regions
from 14 per cent to the residual variance of the entire SDSS main
survey which is estimated, via extrapolation, to be at the 5 per cent
level, see Driver \& Robotham (2010) for details.

\subsubsection{k- and e- corrections}
\begin{figure*}

\centerline{\psfig{file=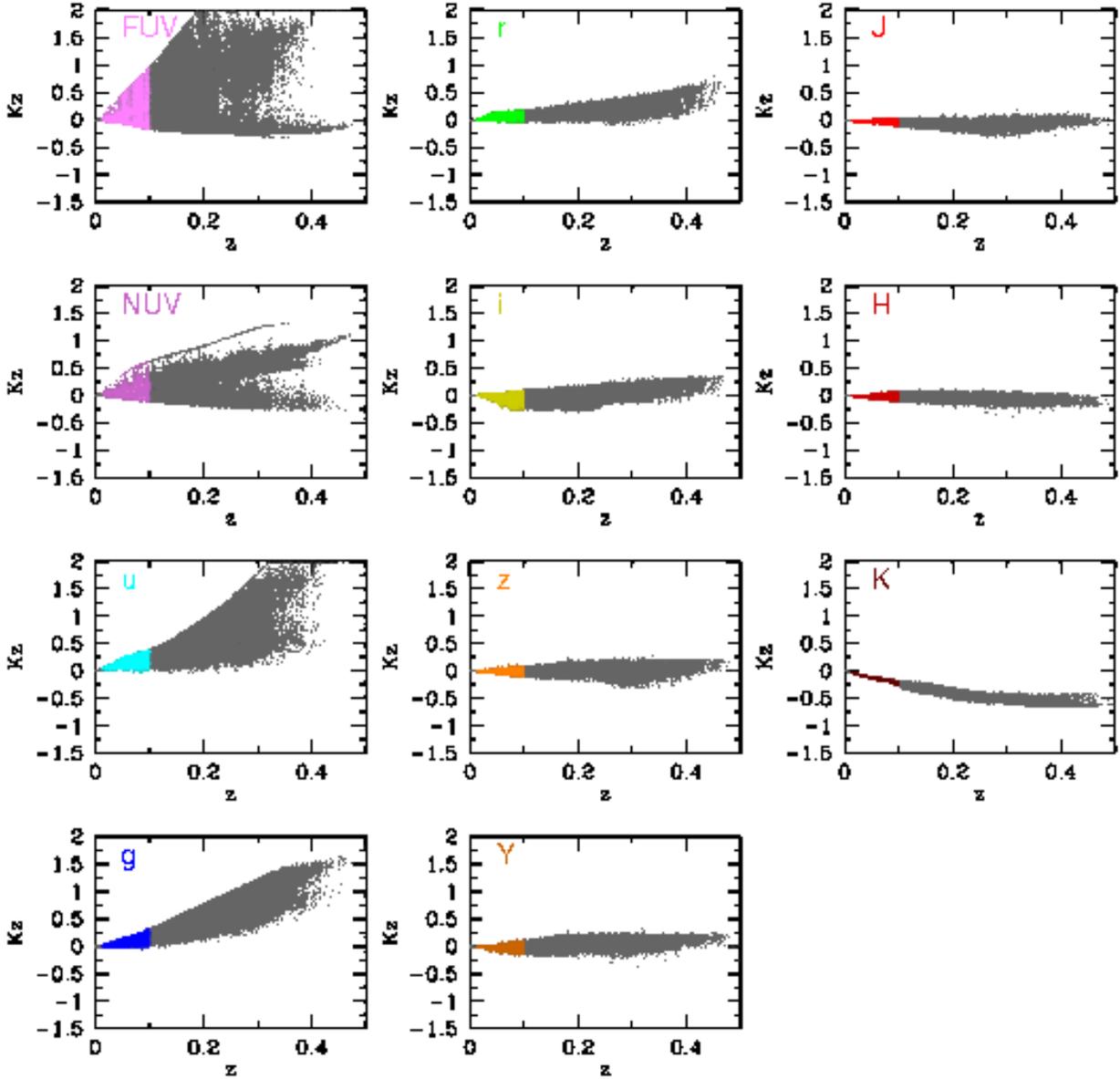,width=\textwidth}}

\caption{\label{fig:kcorr} The k-corrections in each band for the full
  GAMA I sample as derived using KCORRECT (v4.2), and indicating
  generally well behaved distributions. Data in the redshift range
  $0.013 < z < 0.1$ are shown in colour and represent the data used in
  this paper to derive the CSED.}
\end{figure*}

K-corrections are derived for all galaxies using the {\sc KCORRECT}
(v4.2) software of Blanton \& Roweis (2007). We elect to use only the
9 band matched aperture photometry (i.e., SDSS and UKIDSS bands) using
the appropriate SDSS and UKIDSS bandpasses provided with the {\sc
  KCORRECT} software. We then determine the k-corrections in all 11
bands and k-correct to redshift zero. Note that no evolutionary
corrections (e-corrections) are implemented as the redshift range is
low $z<0.1$, this assumption could potentially introduce a small
wavelength bias as the FUV, NUV, u and g bands will be most strongly
affected by any luminosity evolution. Fig.~\ref{fig:kcorr} shows the
k-correction for our sample. Bimodality is clear in the FUV, NUV, and
$u$ bands with values for the FUV becoming quite extreme ($\sim 1$
mag) even at relatively low redshift ($z \sim 0.1$).

\begin{figure*}

\hspace{-9.0cm}{\psfig{file=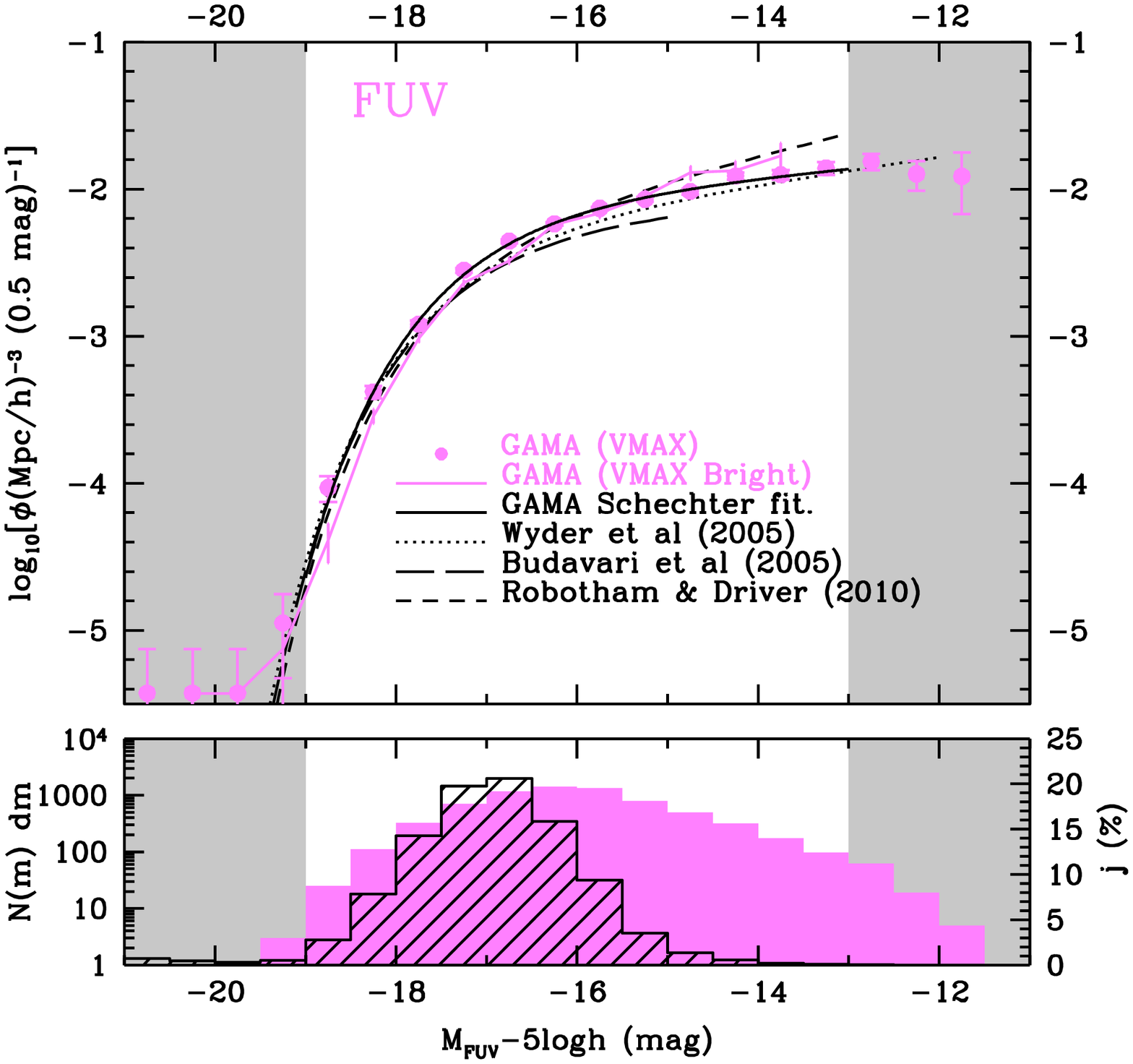,width=\columnwidth}}

\vspace{-8.5cm}

\hspace{9.0cm}{\psfig{file=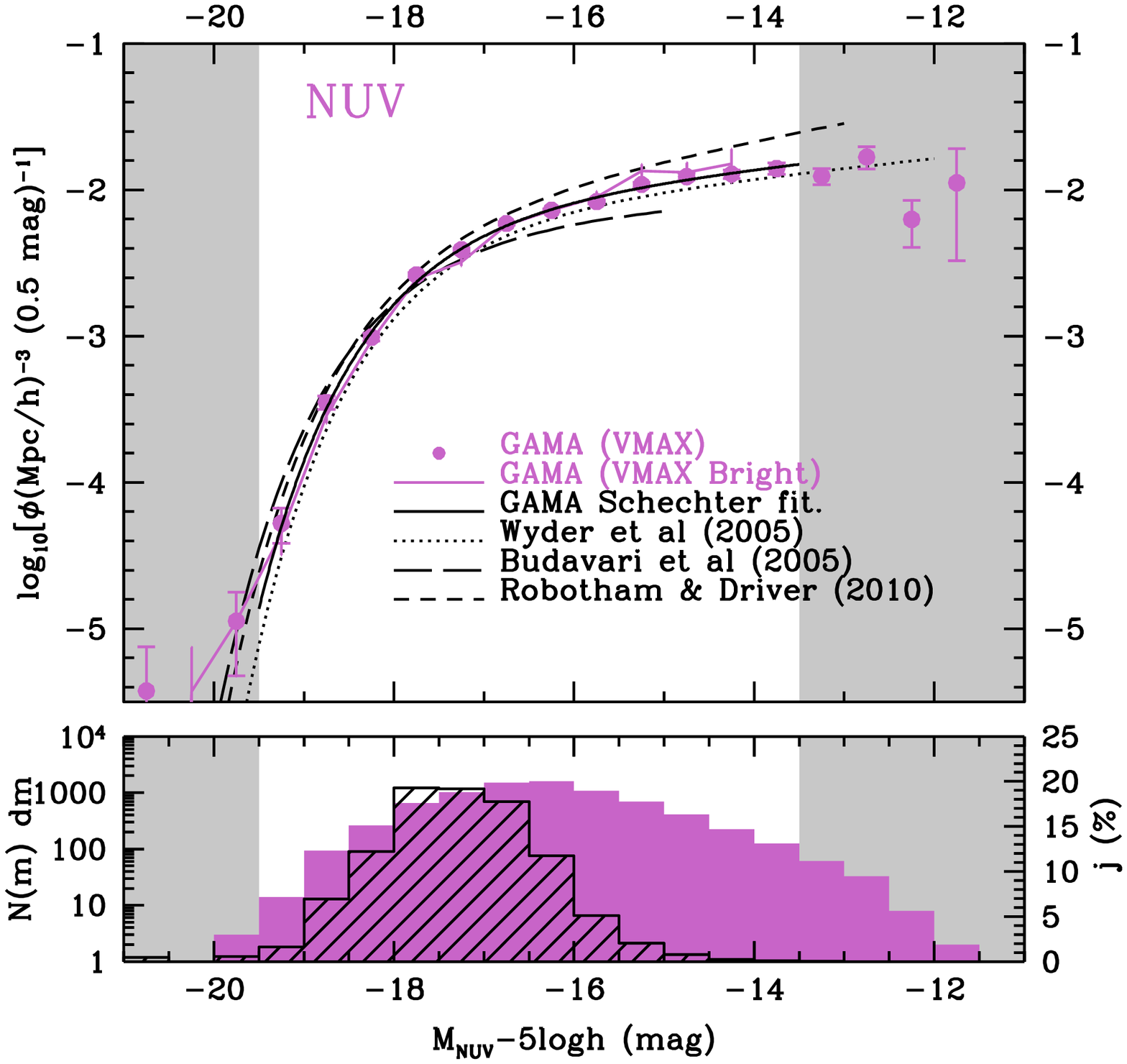,width=\columnwidth}}

\hspace{-9.0cm}{\psfig{file=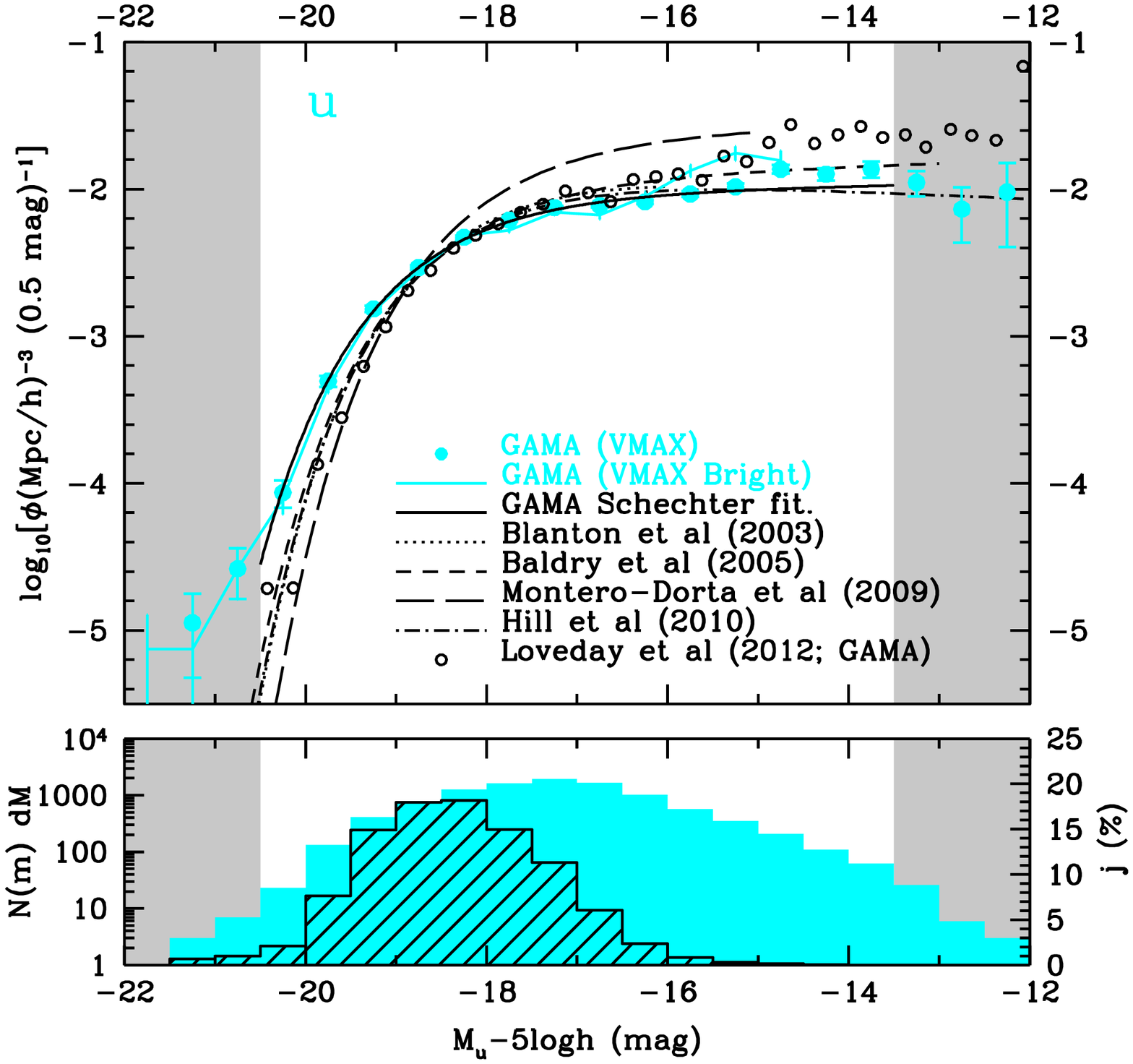,width=\columnwidth}}

\vspace{-8.5cm}

\hspace{9.0cm}{\psfig{file=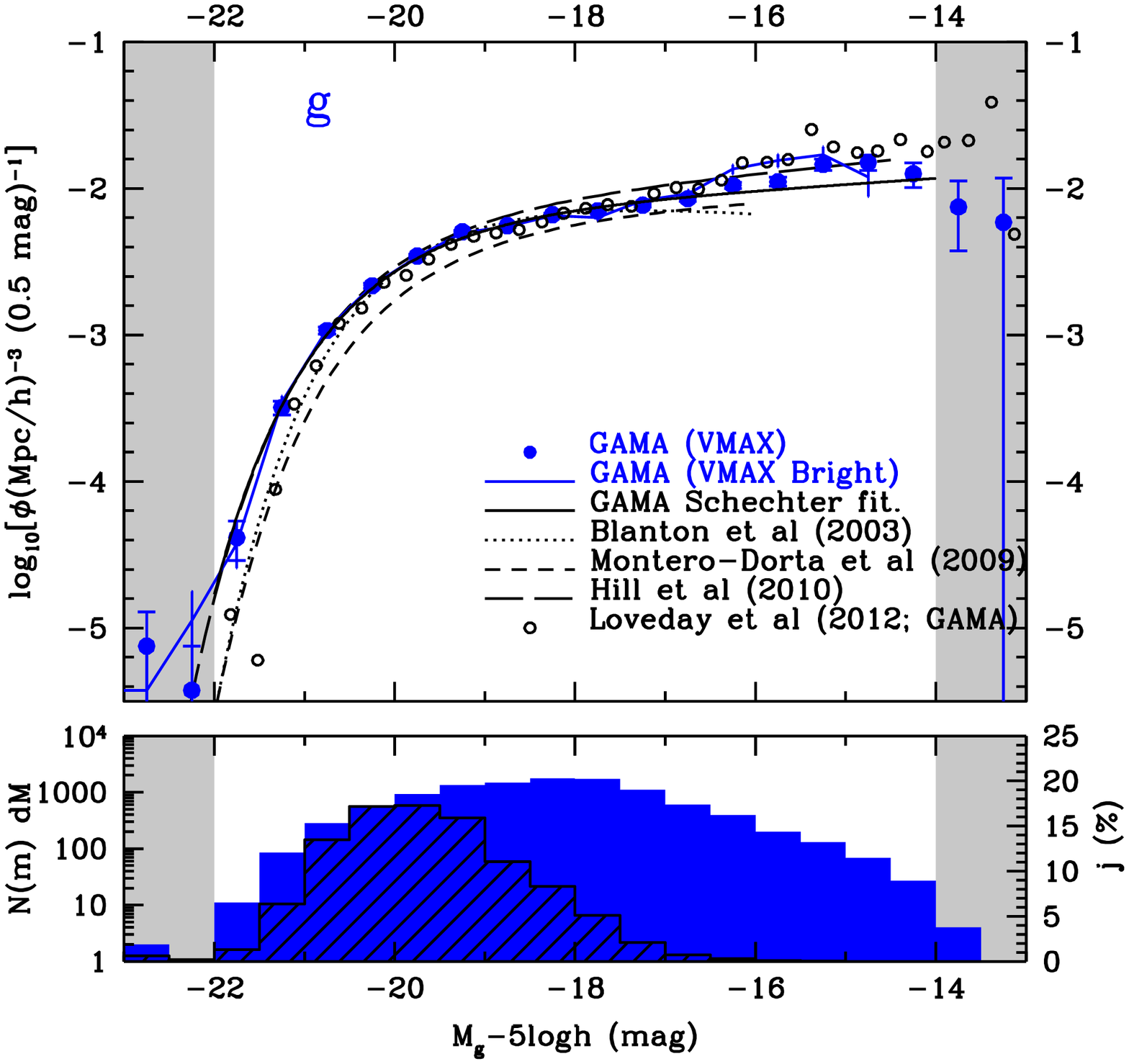,width=\columnwidth}}

\caption{\label{fig:fuvlf} ({\it main panels}) The luminosity
  distribution in the FUV,NUV,$ug$-bands (as indicated) derived via
  $1/V_{\rm Max}$ (solid data points) applying the corrections shown
  in Section 2.5. Where available, pre-existing Schechter function
  fits are shown. The data points with errors use the faint-limits as
  shown in Table~\ref{tab:limits}, col. 4, whereas the line with error
  bars use the more conservative bright limits given in
  Table~\ref{tab:limits}, col. 2. In all cases the faint and bright
  data agree within the errors. ({\it lower panels}) The actual number
  of galaxies used in the derivation of the luminosity distributions
  (solid histogram) and the contribution from each luminosity bin to
  the overall luminosity density (black shaded histogram). The grey
  band at the bright-end indicate as selection boundaries as described
  in the text.}
\end{figure*}

\begin{figure*}

\hspace{-9.0cm}{\psfig{file=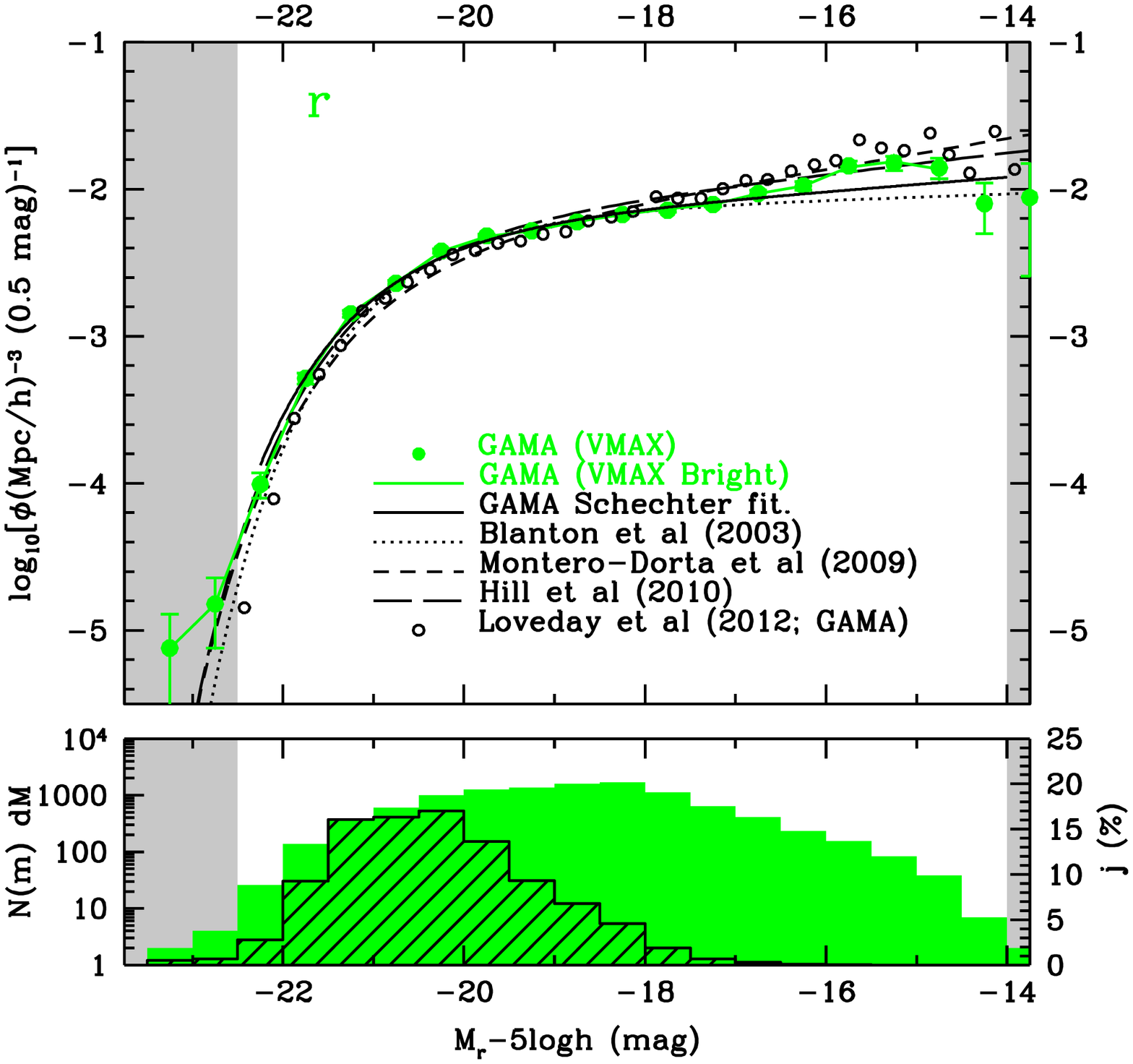,width=\columnwidth}}

\vspace{-8.5cm}

\hspace{9.0cm}{\psfig{file=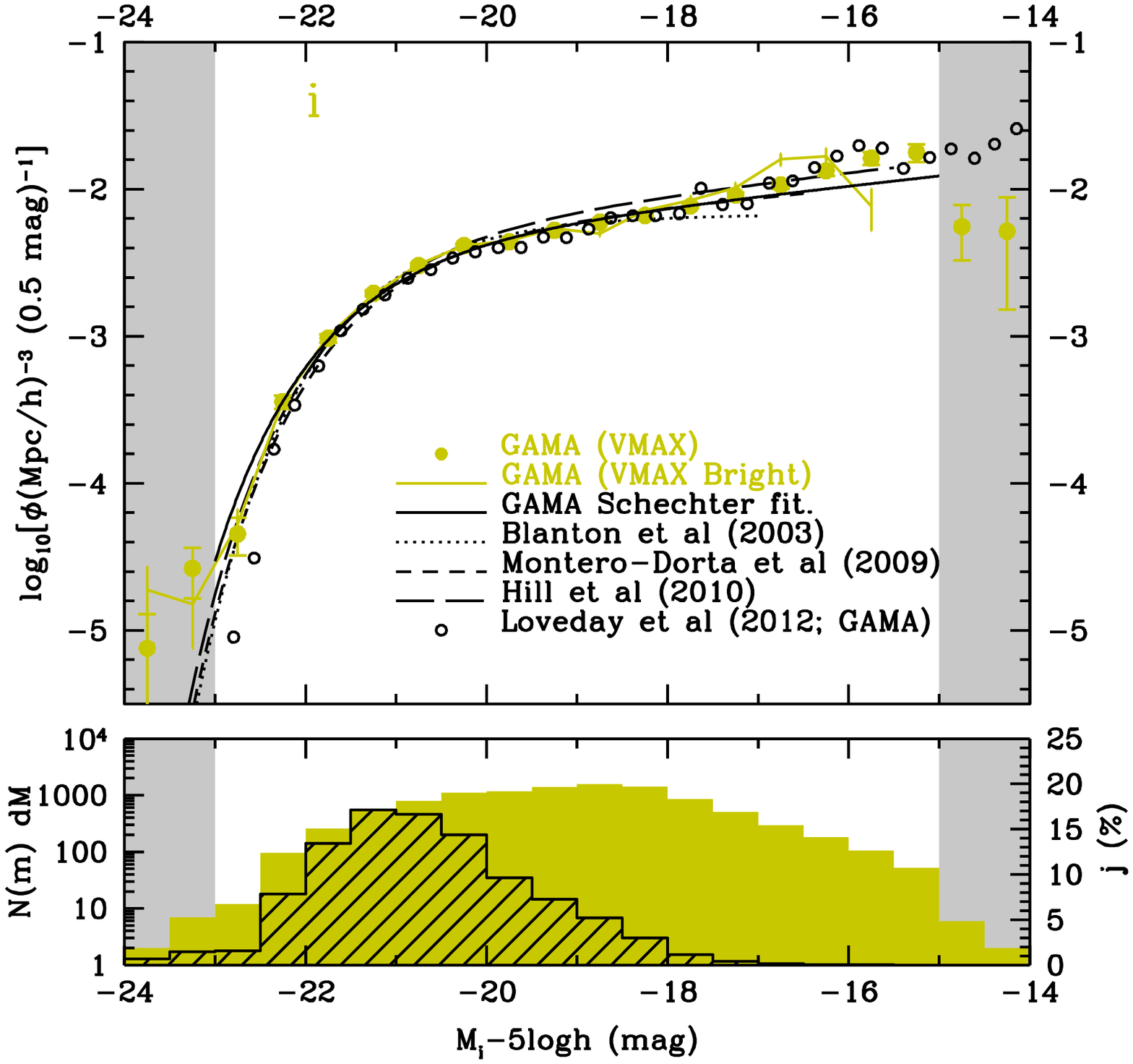,width=\columnwidth}}

\hspace{-9.0cm}{\psfig{file=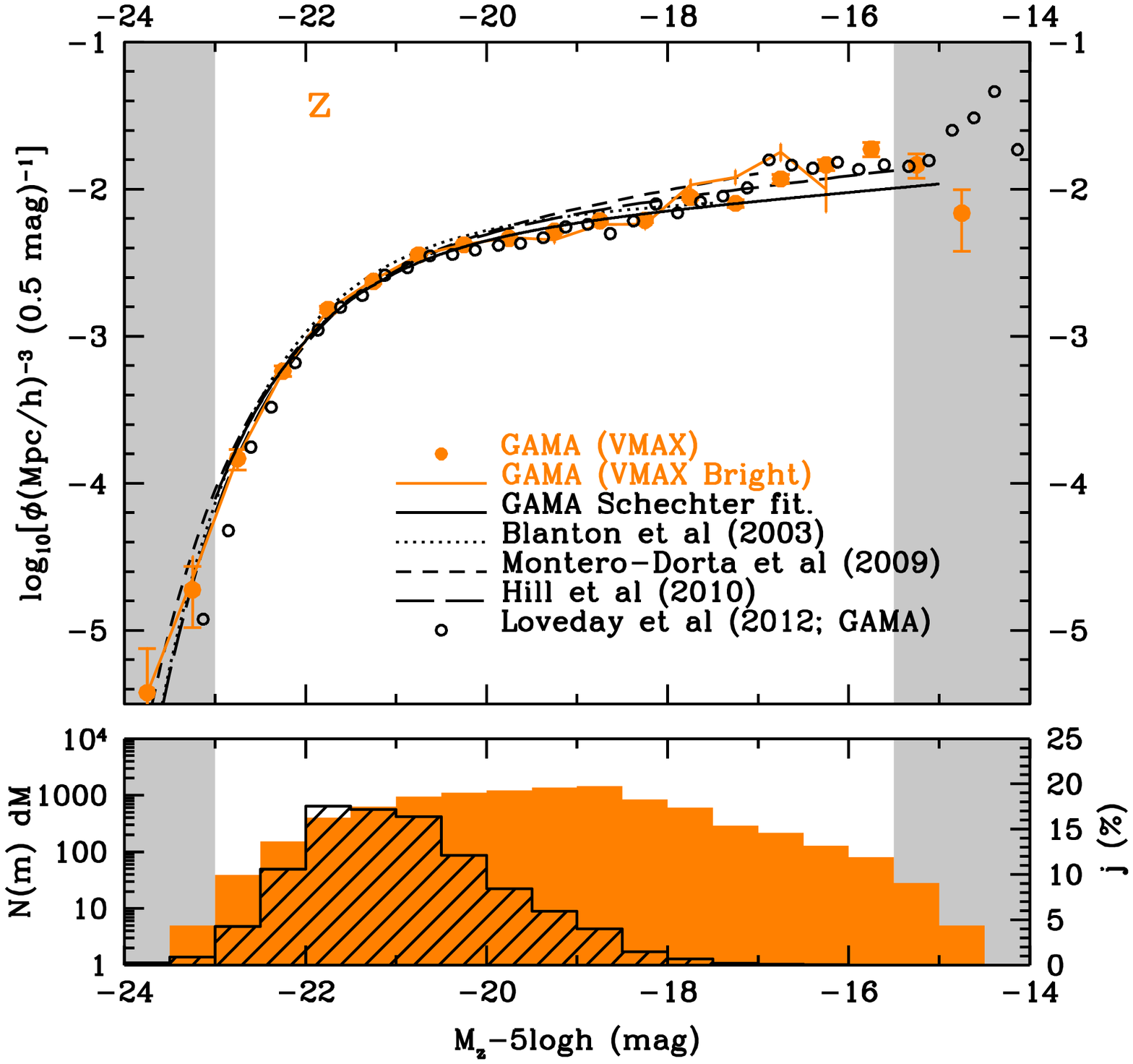,width=\columnwidth}}

\vspace{-8.5cm}

\hspace{9.0cm}{\psfig{file=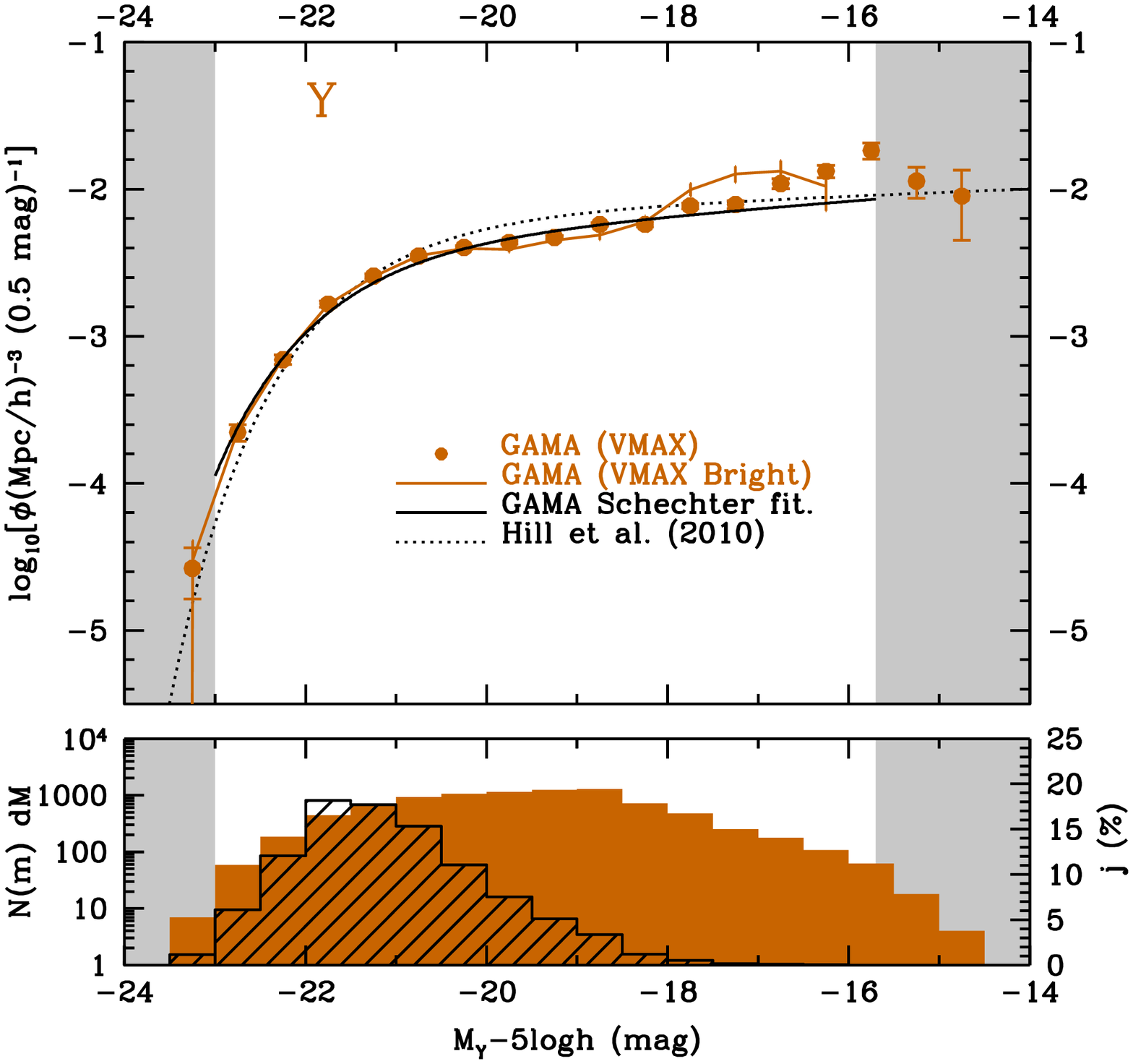,width=\columnwidth}}

\caption{\label{fig:ylf}  As for Fig.~\ref{fig:fuvlf} but in the $rizY$-bands.}
\end{figure*}

\begin{figure*}

\hspace{-9.0cm}{\psfig{file=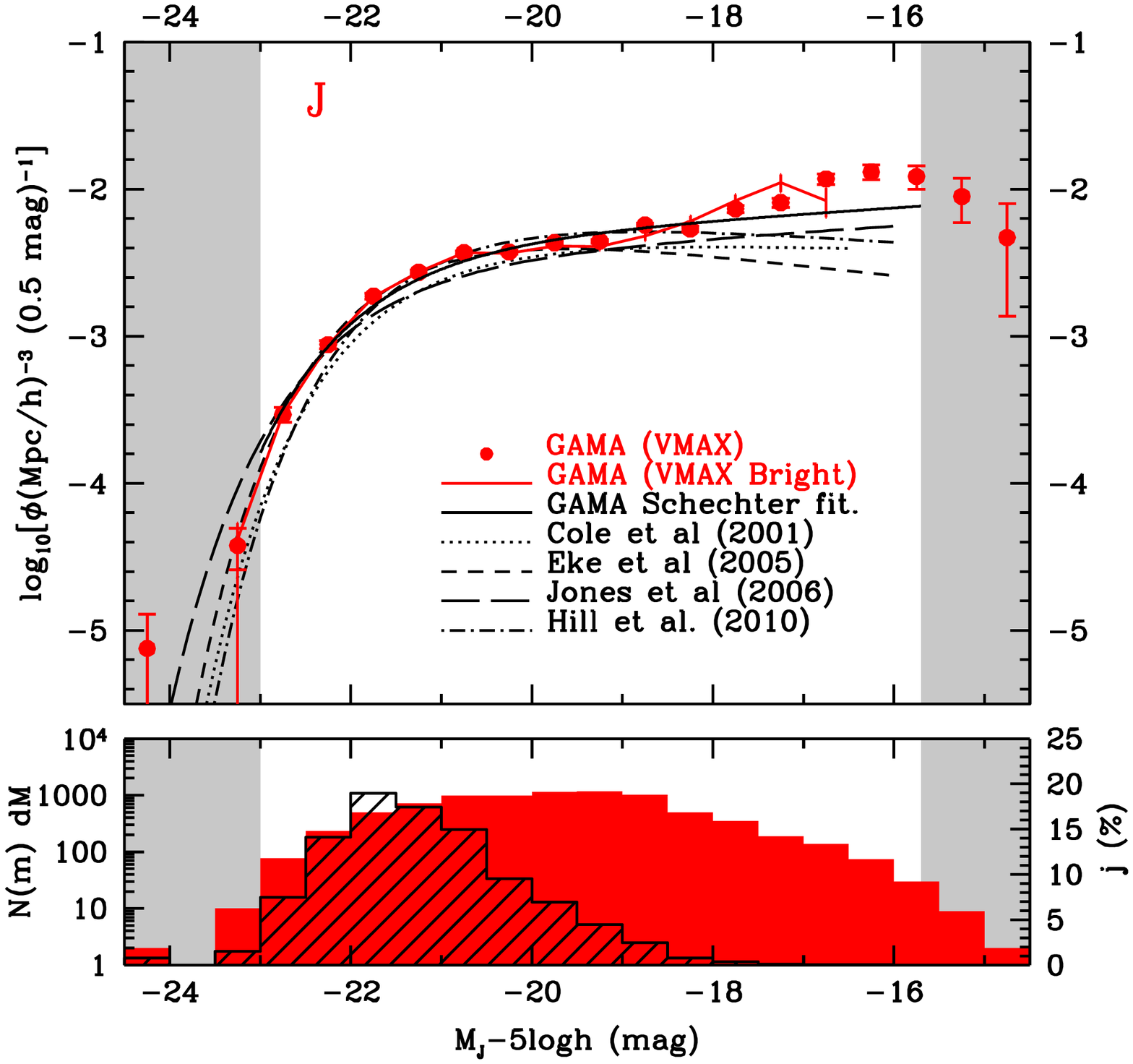,width=\columnwidth}}

\vspace{-8.5cm}

\hspace{9.0cm} {\psfig{file=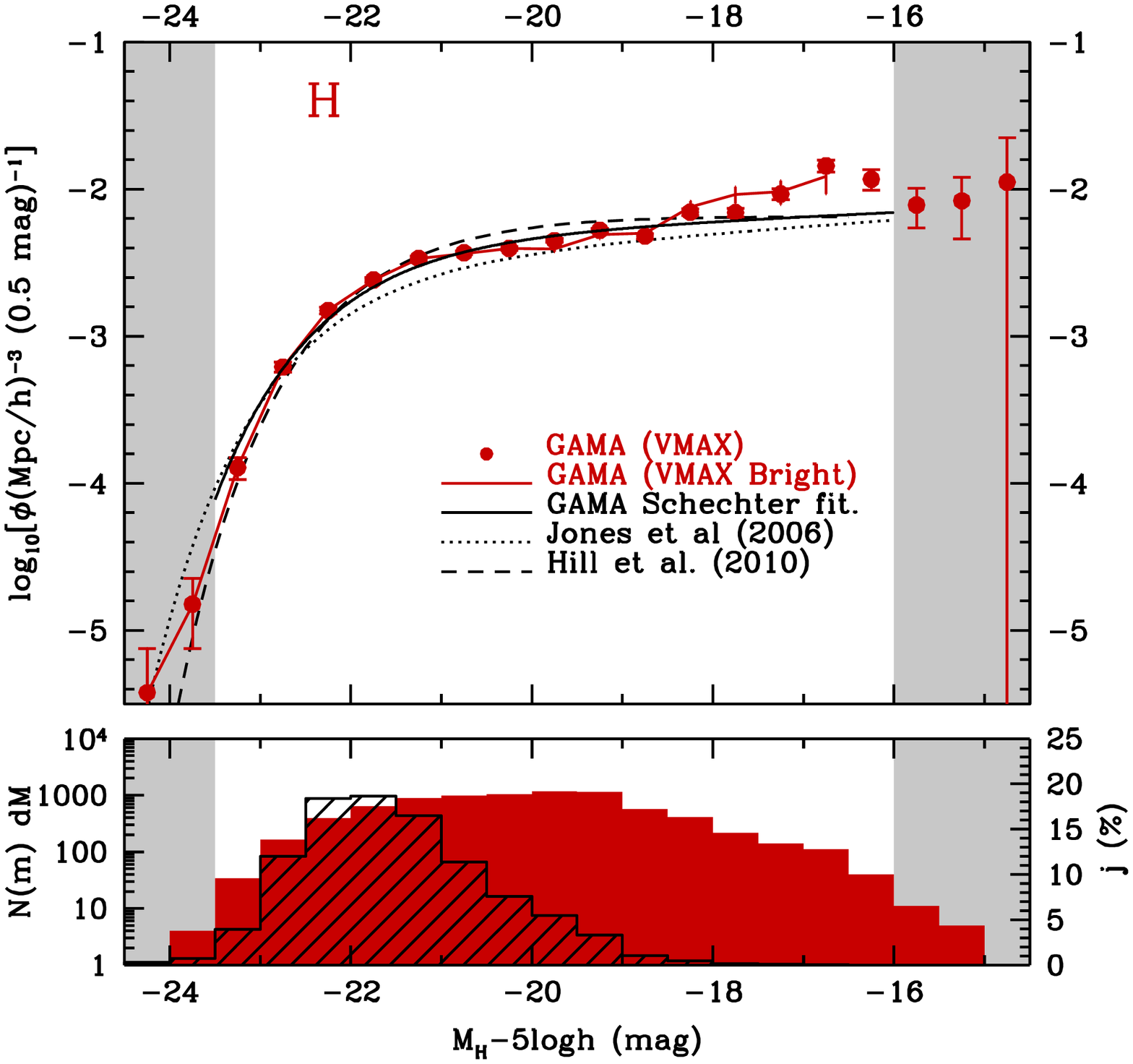,width=\columnwidth}}

{\psfig{file=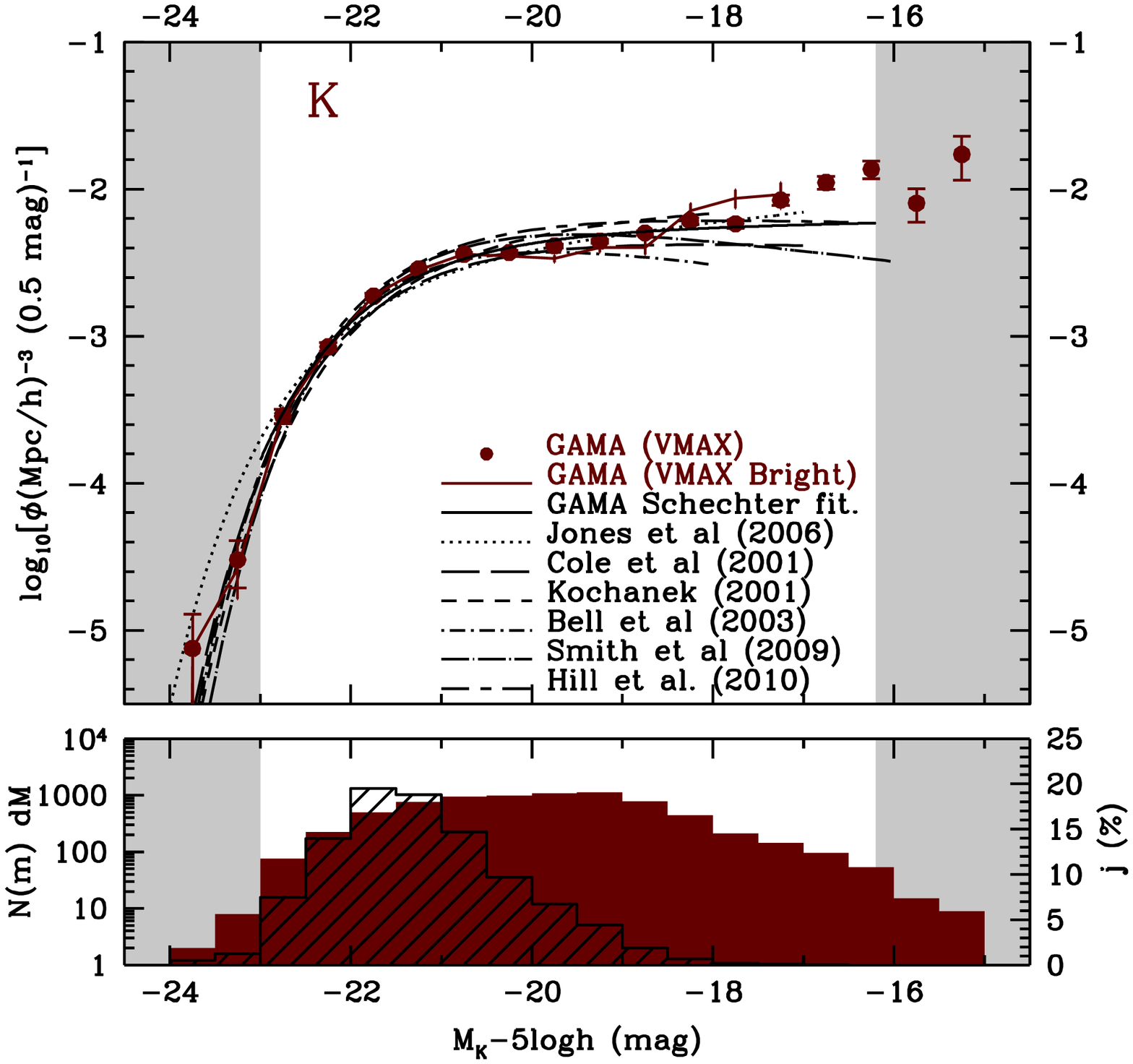,width=\columnwidth}}

\caption{\label{fig:klf}  As for Fig.~\ref{fig:fuvlf} but in the $JHK$-bands.}
\end{figure*}

\subsection{Global luminosity distributions and densities}
The methods of the previous section are now applied to the data
resulting in the output shown in Table~\ref{tab:results}. In all cases
the data are well behaved and an acceptable goodness of fit for the
Schechter function parameters is achieved in all
11-bands. Figs.~\ref{fig:fuvlf}---\ref{fig:klf} (main panels) show our
recovered luminosity distributions, using our $1/V_{\rm Max}$ method
and the bright and faint limits reported in Table~\ref{tab:limits},
along with the Schechter function fit to the $1/V_{\rm Max}$
faint-limit data. Previous results are also plotted, most notably
those from the GALEX, MGC, SDSS and UKIDSS surveys. In $ugriz$ we also
include the recent GAMA $z<0.1$ luminosity functions from Loveday et
al.~(2012) which use the full GAMA area and original SDSS {\it
  Petrosian} photometry, k- and e- shifted to $z=0$. In comparison to
this previous GAMA study we generally see good agreement, although on
close examination there is a consistent offset at the bright-end with
our data shifted brightwards w.r.t Loveday et al.~(2012). We attribute
this to the known difference between Petrosian and Kron photometry for
objects with high S\'ersic index (see for example Graham et al.~2005)
which typically dominate the bright-end.

In comparison to external studies the shape of the curves are mostly
consistent with previous measurements with the greatest spread seen in
the $u$ and $g$ bands (Figs.~\ref{fig:fuvlf}, lower left and right,
respectively). It is important to remember that the GAMA data are, at
the bright-end drawn from a volume limited sample whereas much of the
literature values are flux limited. This can have a significant impact
on the fitted Schechter function parameters as while the values are
unaffected the associated errors will be weighted more uniformly. As a
consequence the Schechter function fits are optimised towards
intermediate absolute magnitudes over purely flux limited
surveys. This subtlety makes it quite tricky to compare Schechter
function parameters directly. However qualitatively the data and fits
shown very good agreement in all bands and over all surveys.

One feature which is distinctly noticeable is the excess (upturns) at
the very faint-end, particularly in the near-IR bands. This has been
previously noted in many papers and explored in more detail for the
GAMA dataset in the $ugriz$ bands by Loveday et al.~(2012).  The
turn-up is most likely brought about by the very red objects at the
very bright-end (i.e., elliptical systems) which essentially create a
plateau shortly below $L^*$ before the more numerous star-forming blue
population with an intrinsically steeper $\alpha$, starts to
dominate. In all cases the figures show the $1/V_{\rm Max}$ results
(which use the limits given in Table~\ref{tab:limits} col, 4) as
datapoints and the $1/V_{\rm Max}$ Bright results (which use the
bright limits given in Table~\ref{tab:limits} col, 2) as a coloured
line. The best fit Schechter function (solid black line) is that
fitted to the $1/V_{\rm Max}$ data points.

\begin{figure}


\centerline{\psfig{file=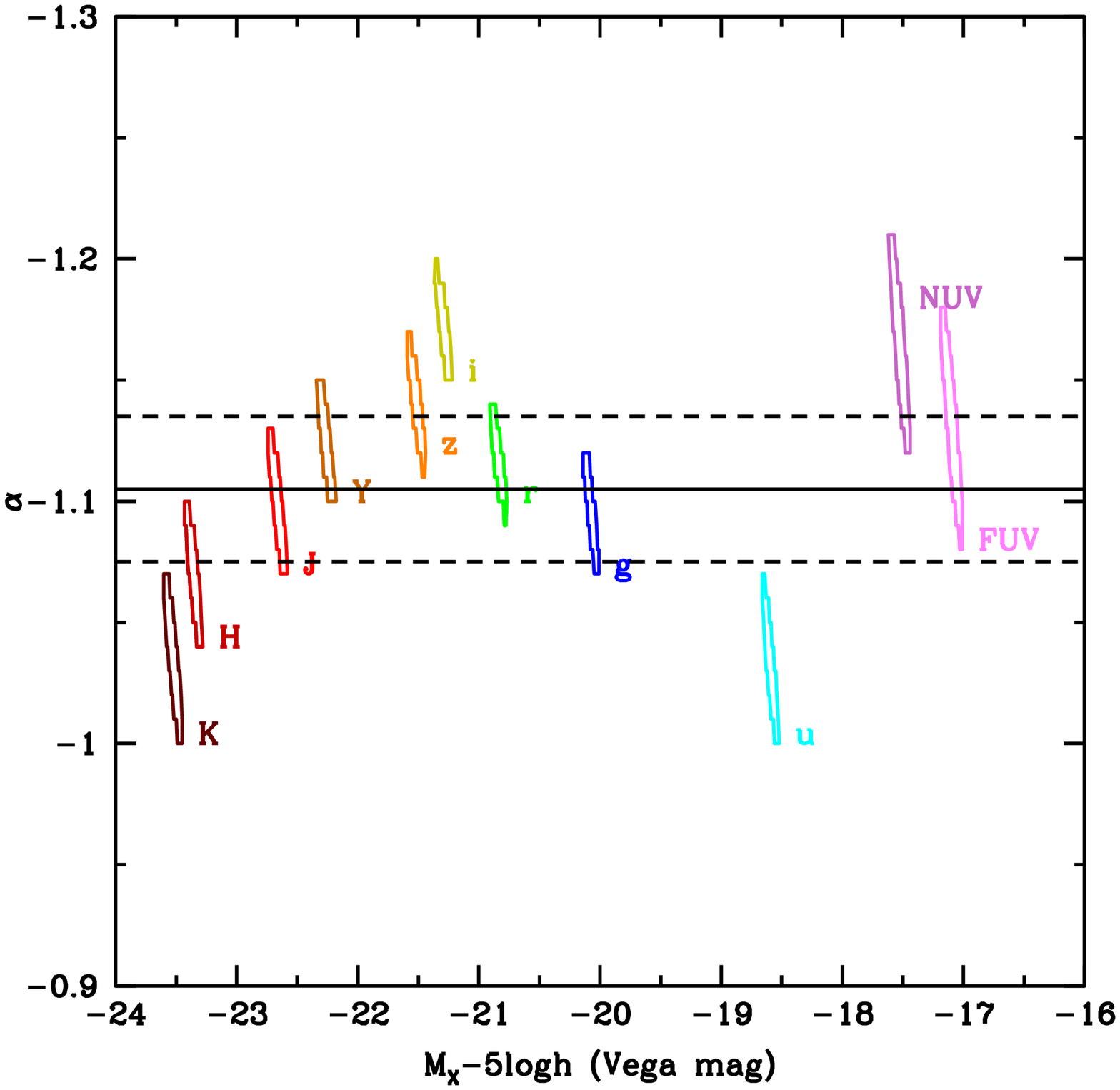,width=\columnwidth}}


\caption{\label{fig:contours} $1-\sigma$ confidence ellipses for the
  Schechter function fits to the data shown in Figs.~\ref{fig:fuvlf}
  --- \ref{fig:klf}.}
\end{figure}

From Fig.~\ref{fig:fuvlf} to \ref{fig:klf} the $1/V_{\rm Max}$ and
$1/V_{\rm Max}$-Bright results agree within the errors, as one would
expect given the significant overlap in the datasets. The lower panels
of Figs.~\ref{fig:fuvlf} to \ref{fig:klf} show the actual number of
galaxies contributing to each bin (solid histogram) and the percentage
contribution of each luminosity bin to the overall luminosity density
(shaded histogram). The grey shading at the bright-end indicates where
fewer than 10 galaxies are contributing to the recovered luminosity
distribution for that bin, and the grey shading at the faint-end
indicates where colour bias will commence (see end of section 3.1)

In this paper we are primarily interested in the integrated luminosity
density rather than the luminosity functions themselves in order to
create the cosmic spectral energy distribution. In all cases we see
that the main contribution to the luminosity density (see shaded
histograms in lower panels of Figs.~\ref{fig:fuvlf} to \ref{fig:klf})
stems from around $L^*$, with a minimal contribution from very bright
and very faint systems. A key concern might be whether there exists a
significant contribution from any low-surface brightness systems not
identified in the original SDSS imaging data. This seems unlikely as
the contribution to the integrated luminosity density falls off at
brighter fluxes than where surface brightness issues are expected to
become significant ($M_r \sim -17$ mag). This confirms the
conclusions made in Driver (1999) and Driver et al.~(2006) that, while
low-surface brightness galaxies may be numerous at very low
luminosities, they contribute a negligible amount to the integrated
luminosity densities at low redshift.

Fig.~\ref{fig:contours} shows the associated $1-\sigma$ error ellipses
for the 11 band Schechter function fits. A faint-end slope parameter
of $\alpha = -1.11 \pm 0.036$ appears to be consistent with all the
error ellipses although some interesting trends are seen with
wavelength.  These trends could be random but could also represent
some faint-end incompleteness in the $u$ and $K$ bands. However one
has to be careful as one typically moves away from the selection
filter ($r$) one samples a narrower range in absolute magnitude and
less into the faint-end upturn which not doubt influences the values
of $\alpha$. However, even at its steepest the relatively flat slope
of $\alpha=-1.11$ implies that relatively little flux density lies
outside the fitted range (as indicated by the shaded histograms in
Figs.~\ref{fig:fuvlf} --- \ref{fig:klf}) and that our luminosity
density estimates should be robust.

\begin{figure}
\centerline{\psfig{file=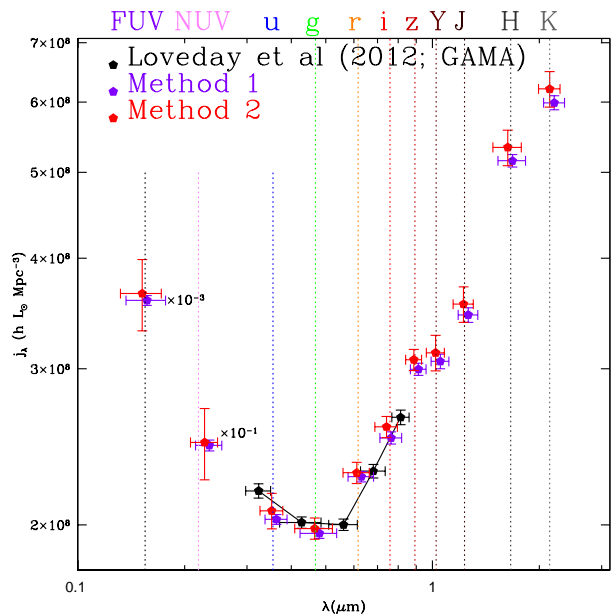,width=\columnwidth}}
\caption{\label{fig:compare} The luminosity density in solar units as
  measured through the 11 bands via the two methods. The FUV and NUV
  data points have been scaled as indicated. The data agree to within
  the errors in all cases. Method 2 is the preferred method now
  carried forward. Note the data have been jittered slightly in
  wavelength for clarity. Also shown are the values taken from Table~6
  (i.e., [${\rm Col.3}+7\times{\rm Col.4}]/8$), of Loveday et
  al.~(2012) derived for the $^{0.1}(ugriz)$ filter set.}
\end{figure}

As described in Section~3 the luminosity density is measured in two
ways and both of these measurements are shown on
Fig.~\ref{fig:compare} and reported in Table~\ref{tab:results}. Also
shown as joined black data points are the luminosity density values
derived by Loveday et al.~(2012) which agree extremely well as one
would expect. Note that these data are shown offset in wavelength as
the values were derived for filters k-corrected to
$z=0.1$. Reassuringly the two measurements from the distinct methods
agree within their quoted errors in all 11-bands implying that there
is no significant error in comparing data derived via alternative
methods. This is because the luminosity distributions are well sampled
around the 'knee' and exhibit relatively modest slopes implying little
contribution to the luminosity density in all bands from the low
luminosity population (as discussed above). Although we have made the
case that the contribution to the luminosity density is well bounded
we acknowledge that the principle caveat to our values is the accuracy
and completeness of the input catalogue which can only be definitively
established via comparison to deeper data. As the GAMA regions will
shortly be surveyed by both VST and VISTA as part of the KIDS and
VIKING ESO Public Survey Programs and we defer a detailed discussion
of the possible effects of incompleteness and photometric error to a
future study.

\subsection{The observed GAMA CSED \label{sec:observedcsed}}
\begin{figure*}
\centerline{\psfig{file=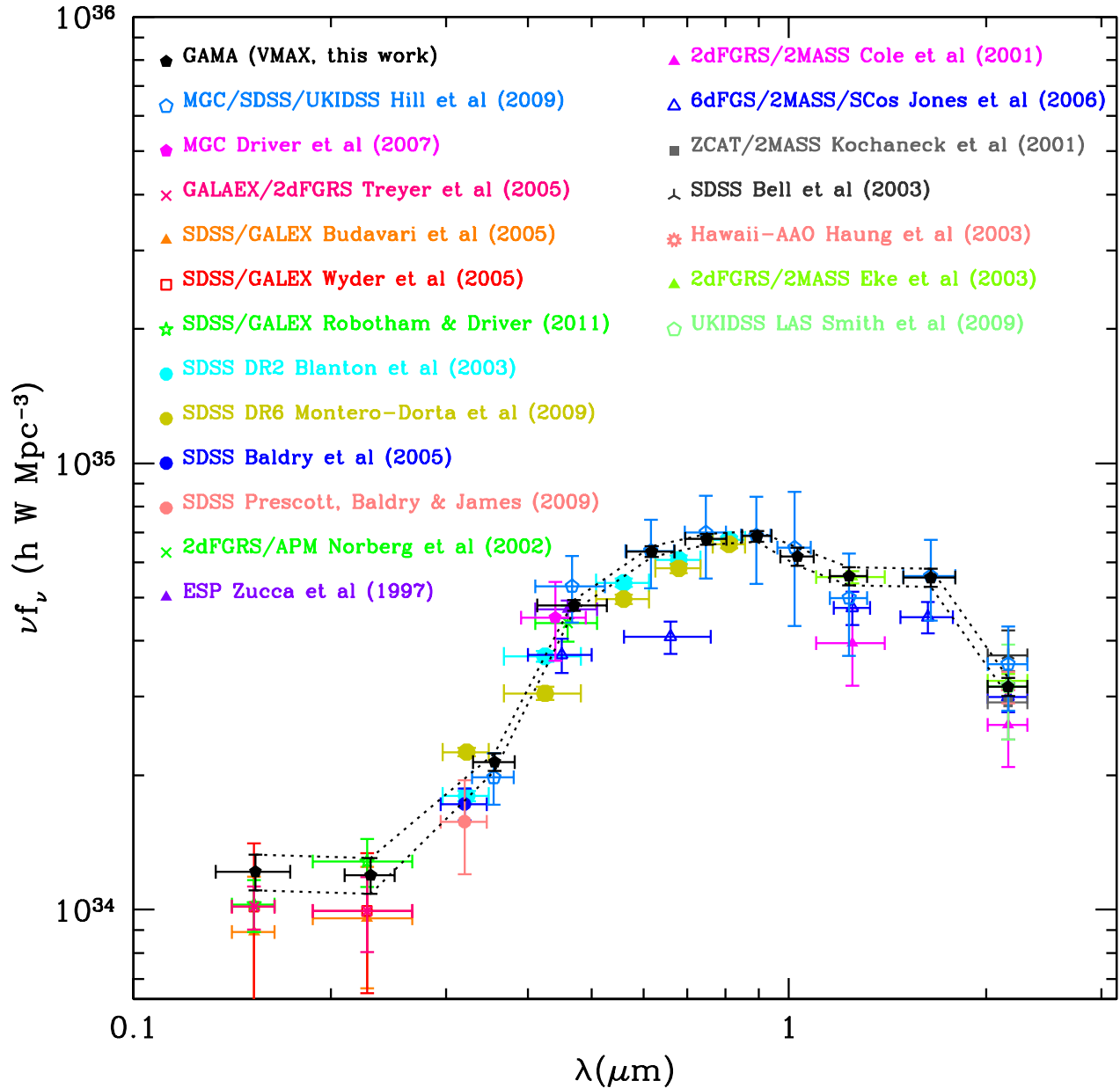,width=\textwidth}}
\caption{\label{fig:csed} The {\it observed} cosmic spectral energy
  distribution from various datasets as indicated. The new GAMA data
  (black squares joined with dotted lines) are overlaid and the
  1-$\sigma$ errors connected via dotted lines which highlight the
  significant improvement in the uncertainty over the previous
  compendium of data.}
\end{figure*}

We now adopt the luminosity density derived from summation of the
individual $1/V_{\rm Max}$ weights for each galaxy, i.e., Method 2.
In Fig.~\ref{fig:csed} we compare these data to previous studies, most
notably Hill et al.~(2010) based on the Millennium Galaxy Catalogue
showing data from $u$ to $K$, Blanton et al.~(2003) and Montero-Dorta
et al.~(2009) which show results from the SDSS from $u$ to $z$, Jones
et al.~(2006) in $b_Jr_FJK$, and Wynder et al.~(2005) and Robotham \&
Driver (2011) in the FUV and NUV. Other typically older datasets are
also shown as indicated on the figure. Note that these data are now
expressed as observed energy densities (i.e., $\epsilon^{\rm Obs}$,
see section 3.3) in which the dependency on the solar SED is divided
out, hence the change in shape from Fig.~\ref{fig:compare} to
Fig.~\ref{fig:csed}. The new GAMA data agree extremely well with
previous studies but show considerably reduced scatter (dotted lines)
across the UV/optical and optical/NIR boundaries. The crucial
improvement is that all the data are drawn from an identical volume
with consistent photometry and therefore robust to wavelength
dependent cosmic variance. In comparison to the previous compendium of
data the GAMA CSED provides at least a factor of 5 improvement and
exhibits a relatively smooth distribution.

\begin{figure}
\centerline{\psfig{file=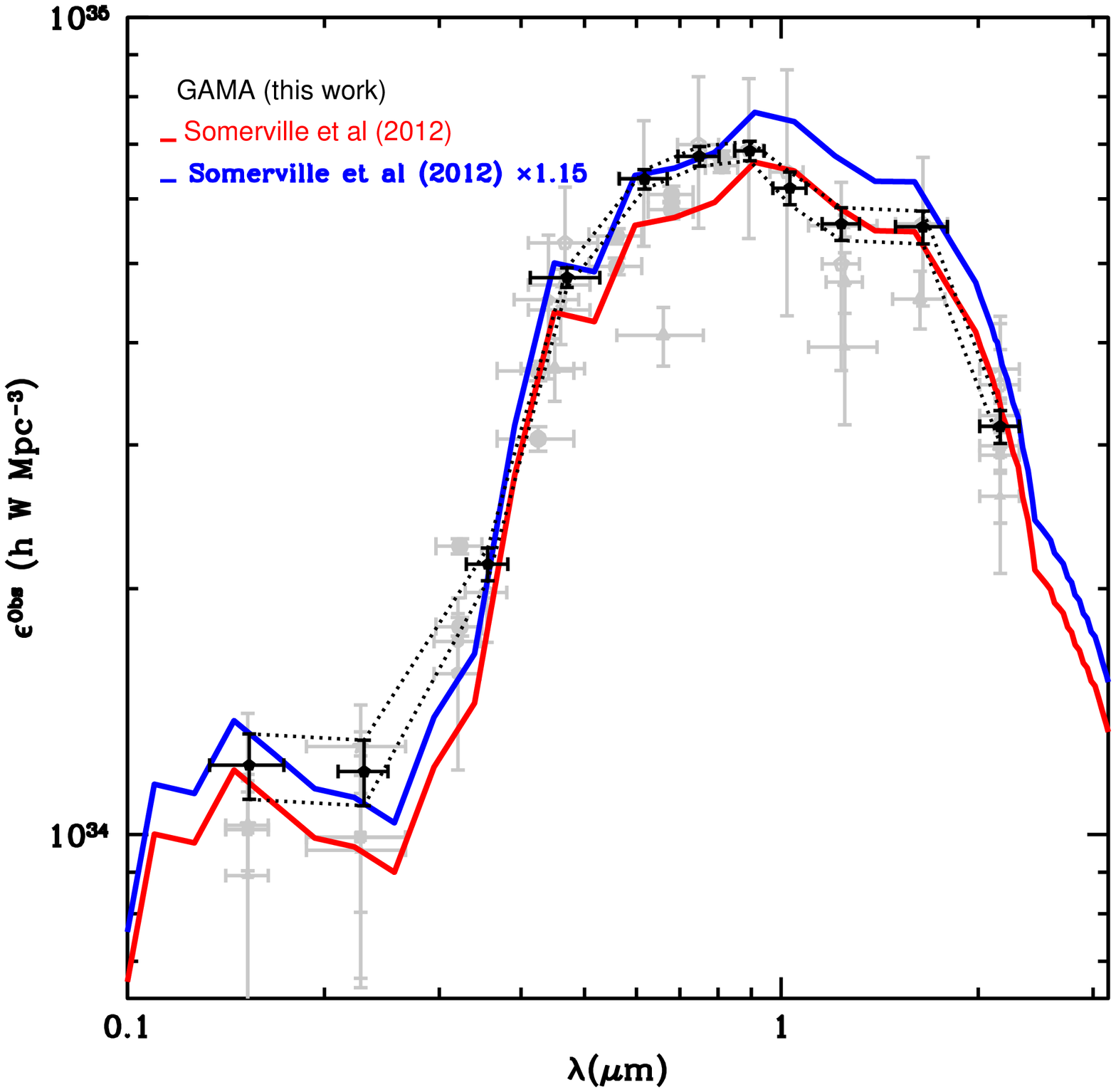,width=\columnwidth}}
\caption{\label{fig:csed2} The {\it observed} cosmic spectral energy
  distribution from the GAMA survey with other data shown in
  grey. Overlaid is the model shown in Fig.~12 of Somerville et
  al.~(2012) unscaled and scaled up by 15 per cent.}
\end{figure}

Perhaps the most noticeable feature in our CSED is the apparent
decline across the transition from the optical to near-IR regime
reminiscent of the discontinuity seen in the earlier study by Baldry
\& Glazebrook~(2003).  Fig.~\ref{fig:csed2} shows the GAMA CSED with
the $z=0$ model from Fig.~12 of Somerville et al.~(2012) overlaid (red
line), and the same model arbitrarily scaled-up by 15 per cent (blue
line). The figure highlights the apparent optical/near-IR
discontinuity with the unscaled model (red line) matching the near-IR
extremely well while the scaled model (blue line) matches the optical
regime very well.  It is difficult to understand the source of this
uncertainty at this time. The two obvious possibilities are a problem
with the data or a problem with the models. The GAMA CSED has been
designed to minimise cosmic variance across the wavelength range by
sampling an identical volume. Similarly great effort has gone into
creating matched aperture photometry from $u$ to $K$ (Hill et
al.~2011) to minimise the photometric uncertainties. It is also clear
that the GAMA LFs are fully consistent with previous measurements,
only a few magnitudes deeper (as indicated by the shaded regions on
Figs.~\ref{fig:fuvlf} to \ref{fig:klf}). In all cases the calculation
of the luminosity density is well defined and Fig.~\ref{fig:compare}
demonstrates that the precise method for measuring the CSED is not
particularly critical with the data generally agreeing within the
errors for both methods (which include methods which extrapolate and
methods which do not). Also the GAMA CSED measurements all lie within
the scatter from the compendium of individual measurements shown in
Fig.~\ref{fig:csed}. Moreover the $Y$ band sits on a linear
interpolation between the $z$ and $J$ bands. Without the $Y$ band data
one would infer a sharp discontinuity between the $z$ and $J$ bands,
however because the $Y$ band perfectly bridges the disjoint it
suggests that the decline is a real physical phenomena.

In terms of modeling, the near-IR is not quite as simple a region as
one might initially expect. Although stable low-mass stars are
expected to dominate the flux, significant contribution to the NIR
flux can also come from the pre-main sequence (shrouded T-Tauri stars
etc) as well as the Thermally Pulsating-Asymptotic Giant Branch
(TP-AGB). In particular significant attention has recently been
focused on the modeling of the TP-AGB which can contribute a
significant amount ($\sim 50$ per cent) of the NIR flux for galaxies
with intermediate aged stellar populations (for a detailed discussions
of this topic see Maraston 2005, 2011 and Henriques et al.~2011). It
is worth noting, that the model constructed in Somerville et
al.~(2012) does not actually include the TP-AGB and were it to be
included the required offset between the red and blue curves would
likely be much greater. The behaviour of the TP-AGB is also know to be
strongly dependent on the metalicity with the progression through the
AGB phase significantly faster for lower-metalicity stars. For
exceptionally low metalicity systems the third dredge-up can be
bypassed entirely, shortening the time over which a TP-AGB star might
contribute significantly to the global SED. We defer a detailed
discussion of this issue but note the suitability of the GAMA data for
either collective or individual SED studies which extend across the
optical/near-IR boundary.

\section{Correcting for dust attenuation}
The GAMA data shown in Fig.~\ref{fig:csed} are all drawn from an
identical volume and therefore robust to cosmic variance as a function
of wavelength. We therefore ascribe the variation between GAMA data
and previous data, in any particular band, as most likely due to
cosmic variance (and in particular variations in the normalisation
estimates of the fitted luminosity functions). The curve and its
integral represents the energy injected into the IGM by the combined
nearby galaxy population, and is therefore a cosmologically
interesting number. However, this energy has been attenuated by the
internal dust distribution within each galaxy. In a series of earlier
papers (Driver et al.~2007; 2008) we quantified the photon escape
fraction for the integrated galaxy population when averaged over all
viewing angles. This was achieved by deriving the galaxy luminosity
function in the $B$ band for samples of varying inclination drawn from
the Millennium Galaxy Catalogue (Liske et al.~2003). The observed
trend of $M^*$ with cos$(i)$ was compared to that predicted by the
sophisticated dust models of Tuffs et al.~(2004, see also Popescu et
al.~2011) and used to constrain the only free parameter, the face on
central opacity, to $\tau_{B}^{f} = 3.8 \pm 0.7$. In Driver et
al.,~(2008) this value was used to predict the average photon escape
fraction as a function of wavelength (0.1 --- 2.1$\mu$m) and the
values adopted are shown in Table~\ref{tab:params}. The errors are
determined from rederiving the average photon escape fraction using
the upper and lower $\tau_{B}^{f}$ values. These corrections are shown
in the final column of Table~\ref{tab:params} and are only applicable
to the disc populations (i.e., Sabc and later).

\begin{figure}

\vspace{-2.0cm}

\centerline{\psfig{file=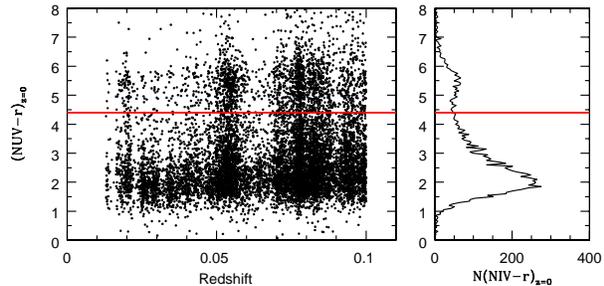,width=\columnwidth}}
\caption{The rest-frame $(NUV-r)$ colour for galaxies with secure redshifts lying in the range $0.013 < z < 0.1$ \label{fig:nuvrcol}}
\end{figure}

In order to accurately correct the CSED for dust attenuation we need
to isolate the elliptical population currently believed to be dust
free.  Rowlands et al.~(2011) recently reported that less than 5\% of
their elliptical sample were directly detected in the far-IR
Herschel-Atlas survey and when far-IR images of the remaining 95 per
cent were stacked the flux recovered implied a mean dust mass of less
than $10^6$ M$_{\odot}$. Hence while not entirely dust free this work
suggests they are certainly between 100 and 1000 $\times$ dust
deficient when compared to similar stellar mass spiral systems. The
approach we take to isolate the ellipticals is informative and
therefore described in full here. Firstly we attempted to isolate the
ellipticals by colour alone. Fig.~\ref{fig:nuvrcol} shows the
rest-frame $(NUV-r)$ colour versus redshift in the range $0.013 < z <
0.1$ which show clear bimodality. Selecting a constant division of
$(NUV-r)_{z=0}=4.4$ mag we repeat the analysis of the previous section
to derive the Schechter function parameters and luminosity
distribution using our $1/V_{\rm Max}$ method.

\begin{figure*}

\centerline{\psfig{file=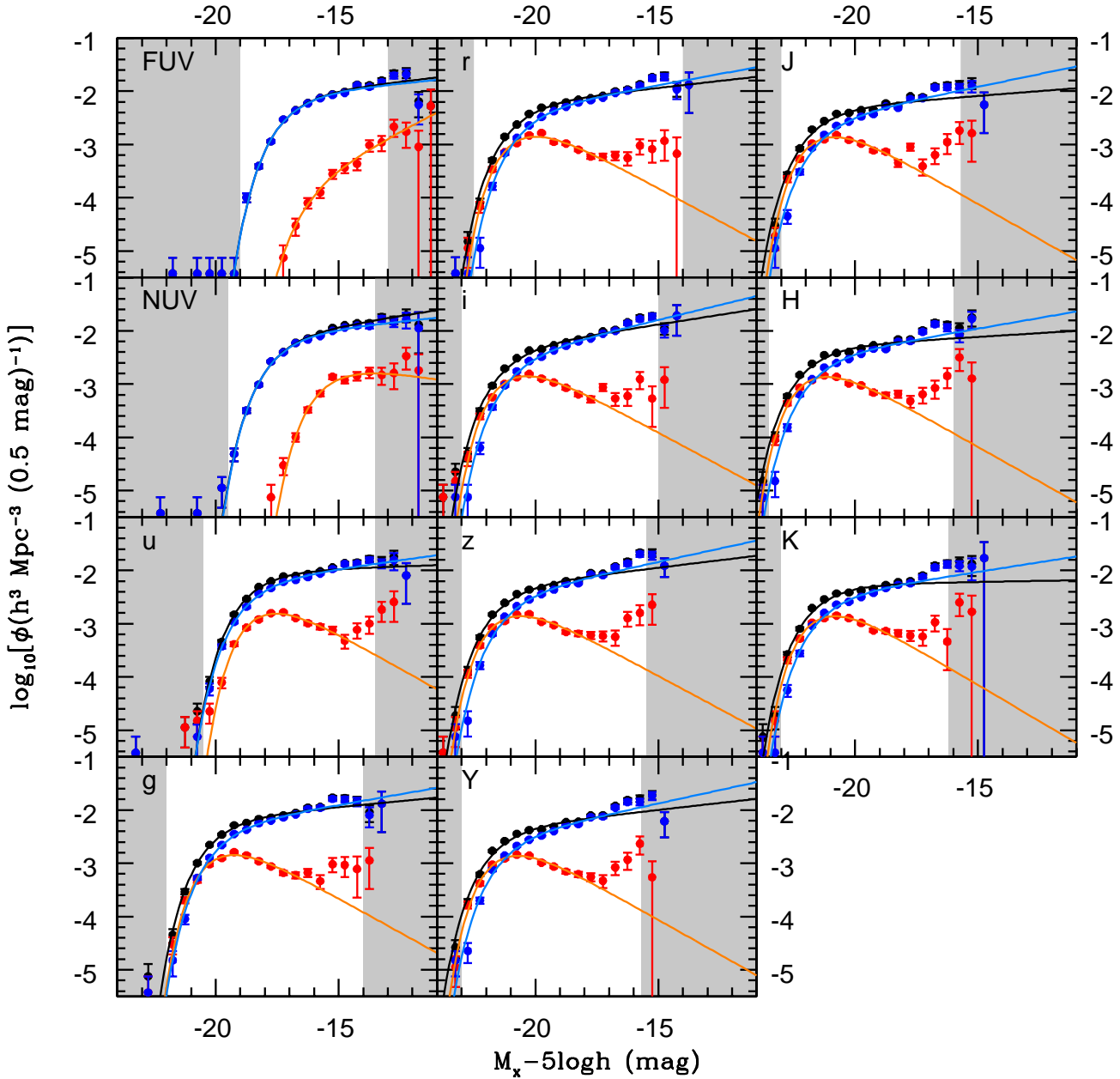,width=\textwidth}}

\caption{\label{fig:allcol} Total luminosity functions and those divided by
  colour across the 11 bands as indicated. The blue data points
  indicate the luminosity distributions and fitted Schechter
  functions for galaxies with $(NUV-r)_{z=0} < 4.4$ and the red data
  with galaxies with $(NUV-r)>4.4$.}

\end{figure*}

\begin{figure}

\centerline{\psfig{file=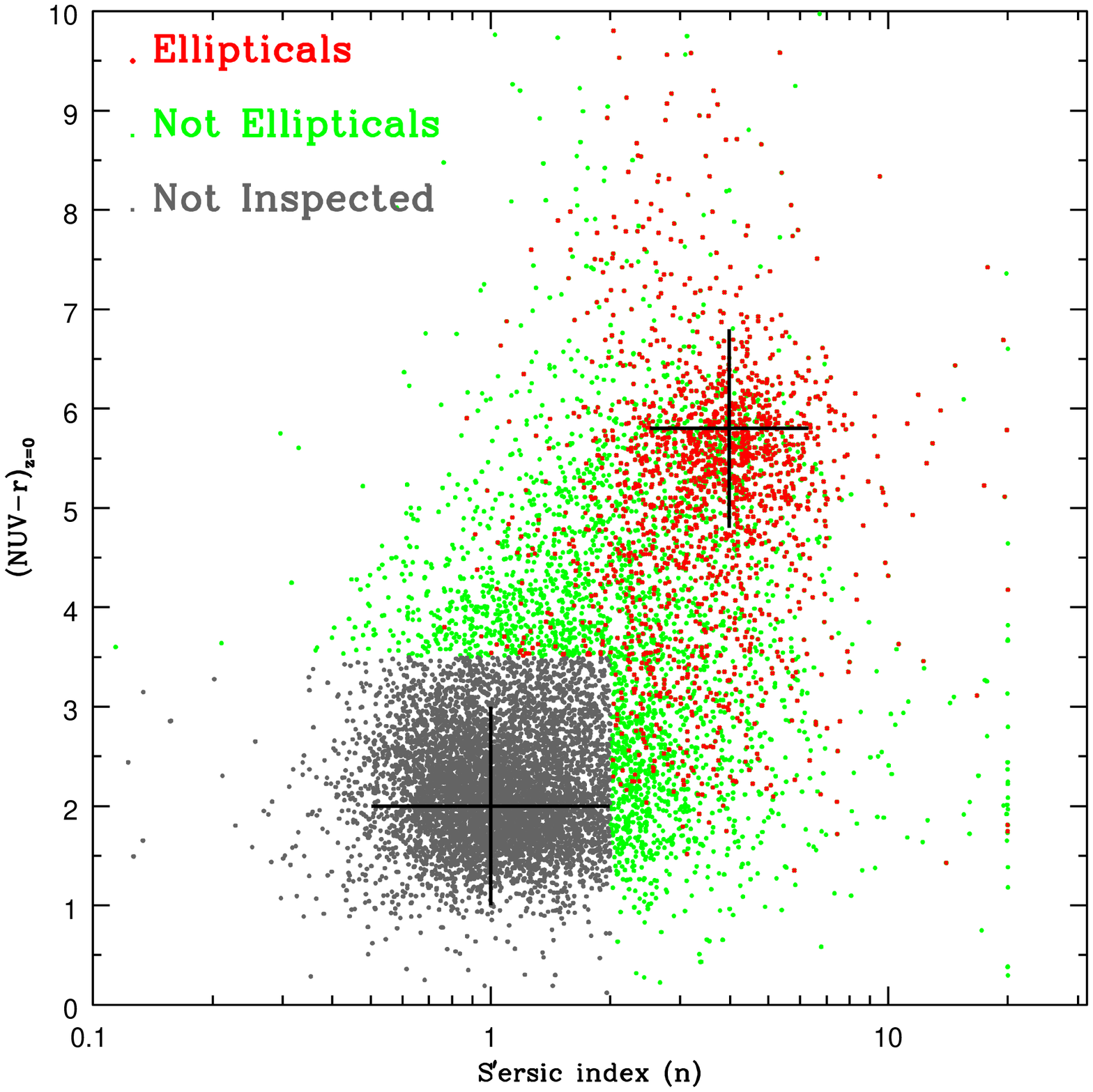,width=\columnwidth}}

\caption{\label{fig:sersic} All galaxies with $n > 2$ or
  $(NUV-r)_{z=0} > 3.5$ have been visually inspected. Those deemed
  ellipticals are indicated by red crosses and those considered not
  elliptical as green dots. The figure indicates that S\'ersic index
  is a better indicator of galaxy type than colour but that for any
  strict automated cut there is serious cross-contamination.}
\end{figure}

The derived luminosity distributions show an interesting effect in
that the red population is clearly bimodal with luminosity, exhibiting
faint-end upturns in most bands. Inspection of the data suggests that
a simple colour cut is overly crude and a significant fraction of
edge-on dusty spirals are being included in the red sample and
responsible for these upturns.  We conclude that colour is not a good
proxy for galaxy type.
In our second attempt we examine the colour-S\'ersic index plane
previously highlighted by Driver et al.~(2005) as a better
discriminator of elliptical systems than colour alone. The S\'ersic
indices are derived via GAMA-SIGMA (Kelvin et al.~2012, an automated
wrapper for GALFIT3, Peng et al.~2010). The fitting process for the
GAMA sample is described in detail in Kelvin et
al.~(2012). Fig.~\ref{fig:sersic} shows the distribution in the
colour-S\'ersic plane for those galaxies with secure redshifts in the
range of interest ($0.013 < z < 0.1$) and flux limited to $r_{\rm
  Kron} < 19.4$ mag. This sample contains 10204 galaxies and exhibits
significant structure. Most noticeable are the two dense cores
indicated by the two black crosses ($n=1,(NUV-r)_{z=0}$=2 and
$n=3.5,(NUV-r)_{z=0}$=5.5) aligned with the anticipated locations of
red elliptical and blue discs.
The two populations clearly overlap and so we resort to a visual
inspection of all systems with $(NUV-r) > 3.5$ mag or $n > 2$. If an
object has no recorded $NUV$ flux it is deemed red and added to the
elliptical/Spheroid-dominated class.  

On Fig.~\ref{fig:sersic} objects classified as
elliptical/Spheroid-dominated, are indicated by a red cross (1,821
systems), those eyeballed but deemed to be not elliptical as green
dots (2,952), and those not inspected as grey dots (5,431). The
criteria used in the visual inspection of the colour postage stamp
images is that a galaxy should be concentrated, smooth, and
ellipsoidal in shape with no indication of internal dust attenuation,
no non-uniformity of colour, and no tidal stream/plume --- all of
which might be indicative of a discy (and therefore dusty)
system. Eyeball classification is by definition subjective however it
is clear from the distribution of the two populations that no
definitive objective measure will separate the
elliptical/Spheroid-dominated (henceforth spheroidal) population
effectively. We do note that a S\'ersic index (vertical) cut would
appear to be more effective than a colour (horizontal) cut alone. This
is primarily because of the colour confusion between old stellar
populations in spheroidal systems and edge-on dust attenuated
spirals. More complex automated strategies to morphologically classify
the GAMA galaxies will be pursued in future papers once the higher
quality imaging data becomes available.

\begin{figure*}

\centerline{\psfig{file=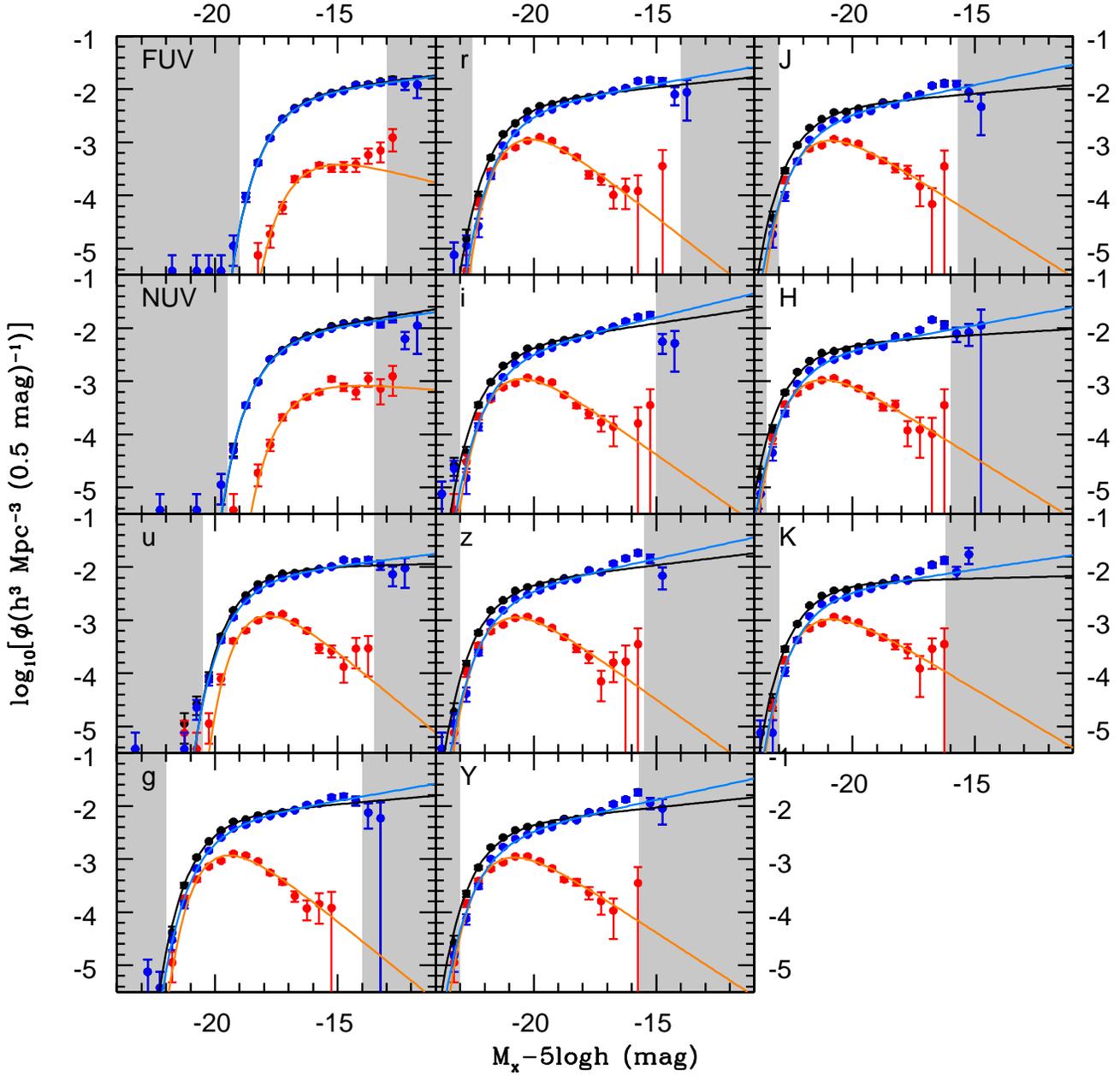,width=\textwidth}}

\caption{\label{fig:allcol2} Total luminosity functions and those divided by
  eye into Spirals (blue points and lines) and ellipticals (red points
  and lines) across the 11 bands as indicated. The eyeball selection
  on a colour-S\'ersic index plane is shown in
  Fig.~\ref{fig:sersic}.}
\end{figure*}

We now re-derive the luminosity functions (Fig.~\ref{fig:allcol2}) in
each band for the spheroidal (red data and lines) assumed to be devoid
of dust, and the remaining populations (blue lines) susceptible to
intrinsic dust attenuation.  Compared to Fig.~\ref{fig:allcol} the
bimodality in the red population is significantly less apparent
suggesting that the eyeball classification is less ambiguous than a
simple colour cut.  Tables~\ref{tab:resultsred} \&
\ref{tab:resultsblue} show the individual Schechter function data for
the spheroidal and discy (non-spheroidal) populations respectively
along with their integrated luminosity densities. We are now in a
position to dust correct the spiral population and sum with the
as-observed spheroidal population to provide the overall CSED
corrected for dust attenuation.

\begin{table*}
\caption{Luminosity function and luminosity density parameters derived
  for each waveband as indicated. \label{tab:results}}
\begin{center}
\begin{tabular}{l|c|c|c|c|c} \hline
Filter  & $M^*-5$logh & $\phi^*$ & $\alpha$ & $J_{\rm Method 1}$ & $J_{\rm Method {\bf 2}}$ \\
        &  (AB mag)      & (10$^{-2}$h$^{3}$Mpc$^{-3}$) & & (10$^{8}L_{\odot}$hMpc$^{-3}$) & (10$^{8}L_{\odot}$hMpc$^{-3}$) \\ \hline
FUV     & $-17.12^{+0.05}_{-0.03}$ & $1.80^{+0.12<}_{-0.07}$ & $-1.14^{+0.03}_{-0.02}$ & $3584^{+27}_{-53}$      & {\bf $3649^{+99}_{-102}$}  \\
NUV     & $-17.54^{+0.04}_{-0.03}$ & $1.77^{+0.08}_{-0.08}$ & $-1.17^{+0.02}_{-0.02}$ & $24.6^{+0.03}_{-0.04}$  & {\bf $24.8^{+0.03}_{-0.04}$}  \\
u       & $-18.60^{+0.03}_{-0.03}$ & $2.03^{+0.05}_{-0.08}$ & $-1.03^{+0.01}_{-0.01}$ & $2.03^{+0.02}_{-0.03}$  & {\bf $2.08^{+0.05}_{-0.06}$}  \\
g       & $-20.09^{+0.03}_{-0.03}$ & $1.47^{+0.04}_{-0.04}$ & $-1.10^{+0.01}_{-0.01}$ & $1.96^{+0.03}_{-0.03}$  & {\bf $1.98^{+0.05}_{-0.06}$}  \\
r       & $-20.86^{+0.04}_{-0.02}$ & $1.24^{+0.04}_{-0.03}$ & $-1.12^{+0.01}_{-0.01}$ & $2.27^{+0.02}_{-0.04}$  & {\bf $2.29^{+0.06}_{-0.06}$}  \\
i       & $-21.30^{+0.03}_{-0.03}$ & $1.00^{+0.04}_{-0.03}$ & $-1.17^{+0.01}_{-0.01}$ & $2.51^{+0.04}_{-0.04}$  & {\bf $2.58^{+0.07}_{-0.07}$}  \\
z       & $-21.52^{+0.03}_{-0.04}$ & $1.02^{+0.03}_{-0.03}$ & $-1.14^{+0.01}_{-0.01}$ & $3.00^{+0.05}_{-0.05}$  & {\bf $3.07^{+0.08}_{-0.09}$}  \\
Y       & $-21.63^{+0.02}_{-0.04}$ & $0.98^{+0.03}_{-0.05}$ & $-1.12^{+0.01}_{-0.02}$ & $3.06^{+0.06}_{-0.05}$  & {\bf $3.13^{+0.08}_{-0.09}$}  \\
J       & $-21.74^{+0.04}_{-0.03}$ & $0.97^{+0.03}_{-0.03}$ & $-1.10^{+0.01}_{-0.01}$ & $3.45^{+0.05}_{-0.08}$  & {\bf $3.55^{+0.10}_{-0.10}$}  \\
H       & $-21.99^{+0.03}_{-0.03}$ & $1.03^{+0.02}_{-0.03}$ & $-1.07^{+0.01}_{-0.01}$ & $5.15^{+0.08}_{-0.08}$  & {\bf $5.33^{+0.15}_{-0.15}$}  \\
K       & $-21.63^{+0.03}_{-0.04}$ & $1.10^{+0.03}_{-0.06}$ & $-1.03^{+0.03}_{-0.06}$ & $6.00^{+0.09}_{-0.13}$  & {\bf $6.21^{+0.17}_{-0.17}$}  \\ 
\end{tabular}
\end{center}
\end{table*}

\begin{table*}
\caption{Luminosity function and luminosity density parameters derived
  for each waveband for the elliptical population as indicated. \label{tab:resultsred}}
\begin{center}
\begin{tabular}{l|c|c|c|c|c} \hline
Filter & $M^*-5$logh & $\phi^*$ & $\alpha$ & $J_{\rm Method 1}$ & $J_{\rm Method {\bf 2}}$ \\
       &  (mag)      & (10$^{-2}$h$^{3}$Mpc$^{-3}$) & & (10$^{8}L_{\odot}$h$Mpc^{-3}$) & (10$^{8}L_{\odot}$h$Mpc^{-3}$) \\ \hline
FUV     & $-16.20^{+0.20}_{-0.20}$ & $0.16^{+0.03}_{-0.03}$ & $-0.70^{+0.21}_{-0.18}$ & $113^{+6}_{^6}$ & {\bf $122^{+3}_{-3}$}  \\
NUV     & $-16.58^{+0.12}_{-0.16}$ & $0.25^{+0.04}_{-0.04}$ & $-0.90^{+0.12}_{-0.11}$ & $1.20^{+0.06}_{^0.06}$  & {\bf $1.25^{+0.03}_{-0.03}$}  \\
u       & $-17.86^{+0.06}_{-0.06}$ & $0.71^{+0.02}_{-0.01}$ & $-0.04^{+0.04}_{-0.06}$ & $0.35^{+0.02}_{-0.01}$  & {\bf $0.37^{+0.01}_{-0.01}$}  \\
g       & $-19.50^{+0.06}_{-0.06}$ & $0.69^{+0.01}_{-0.02}$ & $-0.07^{+0.05}_{-0.04}$ & $0.48^{+0.02}_{-0.02}$  & {\bf $0.50^{+0.01}_{-0.01}$}  \\
r       & $-20.27^{+0.06}_{-0.07}$ & $0.67^{+0.02}_{-0.03}$ & $-0.10^{+0.04}_{-0.05}$ & $0.63^{+0.03}_{-0.03}$  & {\bf $0.66^{+0.02}_{-0.02}$}  \\
i       & $-20.71^{+0.06}_{-0.06}$ & $0.64^{+0.01}_{-0.02}$ & $-0.19^{+0.04}_{-0.04}$ & $0.76^{+0.04}_{-0.02}$  & {\bf $0.78^{+0.02}_{-0.02}$}  \\
z       & $-20.94^{+0.06}_{-0.06}$ & $0.65^{+0.01}_{-0.02}$ & $-0.17^{+0.04}_{-0.04}$ & $0.94^{+0.04}_{-0.04}$  & {\bf $0.97^{+0.03}_{-0.03}$}  \\
Y       & $-21.10^{+0.06}_{-0.06}$ & $0.61^{+0.02}_{-0.02}$ & $-0.25^{+0.04}_{-0.04}$ & $1.00^{+0.04}_{-0.04}$  & {\bf $1.02^{+0.03}_{-0.03}$}  \\
J       & $-21.24^{+0.06}_{-0.05}$ & $0.61^{+0.02}_{-0.02}$ & $-0.27^{+0.04}_{-0.04}$ & $1.17^{+0.04}_{-0.04}$  & {\bf $1.18^{+0.03}_{-0.03}$}  \\
H       & $-21.54^{+0.06}_{-0.05}$ & $0.60^{+0.02}_{-0.01}$ & $-0.28^{+0.04}_{-0.03}$ & $1.74^{+0.06}_{-0.07}$  & {\bf $1.76^{+0.05}_{-0.05}$}  \\
K       & $-21.26^{+0.07}_{-0.06}$ & $0.58^{+0.03}_{-0.01}$ & $-0.31^{+0.05}_{-0.04}$ & $2.02^{+0.07}_{-0.08}$  & {\bf $2.03^{+0.06}_{-0.06}$}  \\ \hline
\end{tabular}
\end{center}
\end{table*}

\begin{table*}
\caption{Luminosity function and luminosity density parameters derived
  for each waveband for the non-elliptical population as
  indicated. \label{tab:resultsblue}}
\begin{center}
\begin{tabular}{l|c|c|c|c|c} \hline
Filter & $M^*-5$logh & $\phi^*$ & $\alpha$ & $J_{\rm Method 1}$ & $J_{\rm Method {\bf 2}}$ \\
       &  (mag)      & (10$^{-2}$h$^{3}$Mpc$^{-3}$) & & (10$^{8}L_{\odot}$h$Mpc^{-3}$) & (10$^{8}L_{\odot}$h$Mpc^{-3}$) \\ \hline
FUV     & $-17.13^{+0.04}_{-0.05}$ & $1.72^{+0.08}_{-0.09}$ & $-1.14^{+0.02}_{-0.01}$ & $3451^{+49}_{-43}$ & {\bf $3527^{+10}_{-10}$}  \\
NUV     & $-17.56^{+0.04}_{-0.04}$ & $1.66^{+0.08}_{-0.08}$ & $-1.16^{+0.02}_{-0.01}$ & $23^{+2}_{-2}$   & {\bf $23.5^{+0.6}_{-0.7}$}  \\
u       & $-18.69^{+0.03}_{-0.04}$ & $1.43^{+0.04}_{-0.07}$ & $-1.14^{+0.01}_{-0.02}$ & $1.68^{+0.03}_{-0.02}$  & {\bf $1.70^{+0.05}_{-0.05}$}  \\
g       & $-20.03^{+0.05}_{-0.02}$ & $1.08^{+0.06}_{-0.03}$ & $-1.20^{+0.02}_{-0.03}$ & $1.48^{+0.02}_{-0.03}$  & {\bf $1.48^{+0.04}_{-0.04}$}  \\
r       & $-20.68^{+0.04}_{-0.02}$ & $0.96^{+0.04}_{-0.02}$ & $-1.20^{+0.01}_{-0.01}$ & $1.61^{+0.02}_{-0.04}$  & {\bf $1.63^{+0.04}_{-0.04}$}  \\
i       & $-21.16^{+0.04}_{-0.04}$ & $0.71^{+0.03}_{-0.03}$ & $-1.28^{+0.01}_{-0.01}$ & $1.75^{+0.03}_{-0.04}$  & {\bf $1.80^{+0.05}_{-0.05}$}  \\
z       & $-21.37^{+0.03}_{-0.04}$ & $0.72^{+0.03}_{-0.04}$ & $-1.25^{+0.01}_{-0.01}$ & $2.05^{+0.04}_{-0.04}$  & {\bf $2.10^{+0.06}_{-0.06}$}  \\
Y       & $-21.54^{+0.04}_{-0.04}$ & $0.63^{+0.03}_{-0.02}$ & $-1.25^{+0.01}_{-0.01}$ & $2.07^{+0.03}_{-0.04}$  & {\bf $2.11^{+0.06}_{-0.06}$}  \\
J       & $-21.67^{+0.05}_{-0.03}$ & $0.60^{+0.04}_{-0.02}$ & $-1.24^{+0.02}_{-0.01}$ & $2.30^{+0.04}_{-0.05}$  & {\bf $2.37^{+0.06}_{-0.06}$}  \\
H       & $-21.92^{+0.04}_{-0.04}$ & $0.65^{+0.03}_{-0.03}$ & $-1.21^{+0.02}_{-0.01}$ & $3.42^{+0.06}_{-0.07}$  & {\bf $3.57^{+0.10}_{-0.10}$}  \\
K       & $-21.55^{+0.05}_{-0.03}$ & $0.70^{+0.04}_{-0.02}$ & $-1.17^{+0.02}_{-0.01}$ & $3.94^{+0.06}_{-0.07}$  & {\bf $4.18^{+0.11}_{-0.12}$}  \\ \hline
\end{tabular}
\end{center}
\end{table*}

\section{The energy budget}
Table~\ref{tab:csed} and Fig.~\ref{fig:energy} show the final cosmic
spectral energy distribution values from FUV through to the $K$ band
both pre- (upper) and post- (lower) dust attenuated. There are two
main motivations for having constructed these data. The first is to
provide an estimate of the energy production rate in the Universe
today pre dust attenuation, and the second is to constrain models of
galaxy formation. We defer a detailed discussion of the latter to a
companion paper which introduces the two-phase galaxy formation
model (Driver et al.~2012). However before calculating the pre- and
post-corrected energy density we first digress to provide an
independent estimate of the the local star-formation rate from our
dust-corrected FUV luminosity density.

\subsection{The local star-formation rate}
The dust corrected FUV luminosity density is recognised as a good
proxy for the star-formation rate at the median redshift of the study
concerned. This is because typically only massive stars with lifetimes
less than 10Myrs contribute significant FUV flux. Following Robotham
\& Driver (2011) we use the standard Kennicutt (1998) prescription
based on a Salpeter (1955) initial mass-function, whereby
$SFR_{FUV}(M_{\odot} $yr$^{-1}) = 1.4 \times 10^{-28} L_{\nu} ($ergs$
s^{-1} $Hz$^{-1})$ or $1.4 \times 10^{-28} \frac{\epsilon^{\rm Int}
  \lambda_p}{10^{-7}c}$ to derive a star-formation rate at a volume
weighted redshift of $z=0.078$ of 
$0.034\pm 0.003$(Random)$ \pm
0.009$(Systematic, Dust) $\pm 0.002$(Systematic, cosmic variance) h
M$_{\odot}$ yr$^{-1}$ Mpc$^{-3}$. 
This is consistent with the
compendium of results shown in Table~4 of Robotham \& Driver (2011)
and only $\sim 9$ per cent higher than the most recent values derived
by Wyder et al.~(2005) and Robotham \& Driver (2011) of $0.0311 \pm
0.006$ and $0.0312 \pm 0.002$ h M$_{\odot}$ yr$^{-1}$ Mpc$^{-3}$
respectively. Essentially all the star-formation is in the
non-spheroidal systems and although formally there is $2.7 \times
10^{-4}$ h M$_{\odot}$ yr$^{-1}$Mpc$^{-3}$ in the spheroid population it
is highly likely that this flux might be dominated by a small number
of blue spheroids (see Fig.~\ref{fig:sersic}).

\subsection{The instantaneous energy output of the nearby Universe}
Of cosmological significance is the total amount of starlight being
produced in the local Universe at the present epoch and the amount
which escapes into the IGM and is detectable in UV-NIR bandpasses. The
discrepancy between the energy generated and that which escapes into
the IGM in the optical/near-IR window, must equate to the local far-IR
dust emission if starlight is the only source of heating. The
implicit assumption being that the missing light is being attenuated
by dust. 

To evaluate the energy within the pre-attenuated CSED we must identify
a suitable fitting function with which to represent the data, and will
enable an extrapolation over the full UV to mid-IR regime. To do this
we adopt the predicted spheroid and disc CSEDs from the zero-free
parameter two-phase galaxy formation model of Driver et al.~(2012) and
renormalise them slightly to better fit the data. The renormalised
model CSEDs are shown on Fig.~\ref{fig:energy} (lower panel) as the red and
blue curve (for spheroids and discs), and provide a very good match to
the data (although note once again the difficulty in matching the
sharp decline in the CSED at the optical/near-IR boundary, see
discussion in Section~\ref{sec:observedcsed}). For full details of
these models see Driver et al.~(2012) but in brief they adopt a
distinct cosmic star-formation history for spheroids and discs, an
evolving metalicity, the PEGASE star-formation code, and a Baldry \&
Glazebrook (2003) initial mass function.

We now integrate the models which includes extrapolation to the Lyman
limit. We find a total intrinsic energy density of: $(1.8 \pm 0.3)
\times 10^{35}$ W h Mpc$^{-3}$. This is subdivided into an energy
budget of $(1.45 \pm 0.2) \times 10^{35}$ and $(0.8 \pm 0.1) \times
10^{35}$ W h Mpc$^{-3}$ for the non-spheroid population before and
after attenuation, and $(3.6 \pm 0.5) \times 10^{34}$ W h Mpc$^{-3}$
for the Spheroid population. The final errors in the dust corrected
data are almost entirely dominated by the uncertainty in the photon
escape fraction (see Table~\ref{tab:params}, col 5). Our new local
($<z> \sim 0.078$) energy production values ($\epsilon^{\rm Int}$) can
be compared to our earlier estimate in Driver et al.~(2008) of
$\epsilon^{\rm Int}=(1.6 \pm 0.2) \times 10^{35}$ W h
Mpc$^{-3}$. These integrated energy values also agree within the
errors to the fairly broad ranges deduced in Baldry \& Glazebrook
(2003) of $(1.2$ --- $1.7) \times 10^{35}$ W h Mpc$^{-3}$ for
attenuated starlight of which $(0.3$---$0.7) \times 10^{35}$ W h
Mpc$^{-3}$ is reprocessed by dust.

\subsection{Predicting the local far-IR emission}
\begin{figure*}

\centerline{\psfig{file=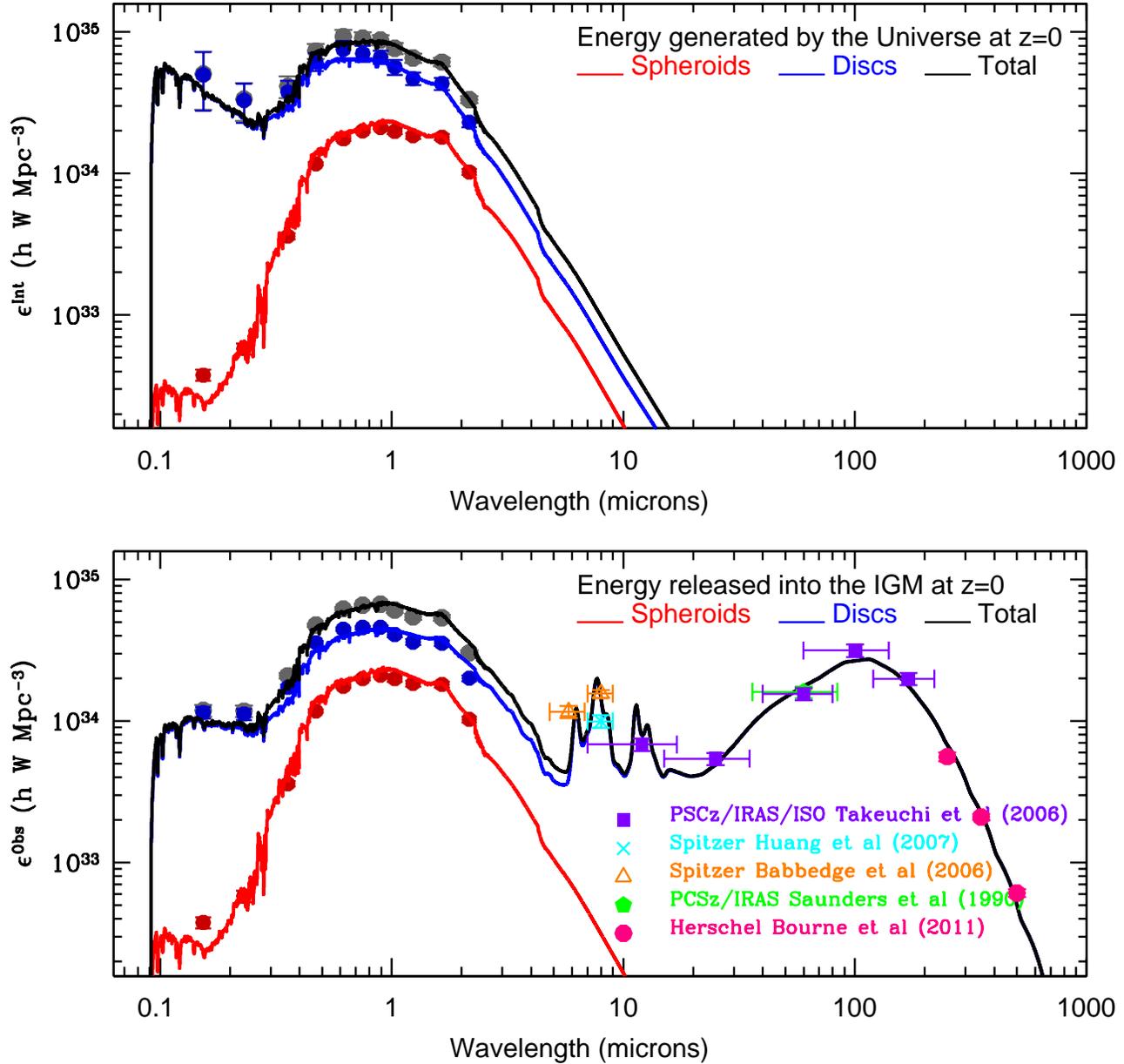,width=\textwidth}}

\caption{\label{fig:energy} The energy output of the Universe from UV to FIR from spheroid dominated (red) and non-spheroid systems (blue) pre- (upper) and post- (lower) dust attenuation.}
\end{figure*}

The difference between the total post and pre- attenuated energies is
presumed to be re-radiated in the far-IR by dust grains and equates to
a total energy of $(6 \pm 1) \times 10^{34}$ W h Mpc$^{-3}$ at
$z<0.1$, i.e., ($35 \pm 3$) per cent of the energy produced by stars
is extracted by dust and reradiated in the far-IR (slightly lower than
our previous estimate of ($43 \pm 5$) per cent; Driver et al.~2008).
Assuming that the attenuated starlight is entirely absorbed by dust
and dominates over all other dust heating processes we can use this
value to predict the far-IR CSED. To do this we adopt the average of
the Dale \& Helou (2002) models 34 and 42, following Baldry \&
Glazebrook (2003) and Driver et al.~(2008). We then renormalise the
dust emission curve until it contains the same amount of energy as
that lost to the attenuation of stellar light. The attenuated
starlight CSED and dust emission CSED can then be added to provide a
complete description of the CSED from the Lyman limit (0.1 $\mu$m) to
the sub-mm ($<1$mm) wavebands. This wavelength regime is crucial as it
entirely dominates the energy production budget of the nearby
Universe. The full 0.1$\mu$m---1mm CSED is shown in the lower panel of
Fig.~\ref{fig:energy} (black curve) and represents a prediction of the
full CSED based on the GAMA UV/optical/near-IR data coupled with the
photon escape fractions given in Driver et al.~(2008).

On Table~\ref{tab:farir} we show how our prediction compares to the
currently available data from various mid and far-IR studies. For
completeness the actual values and their sources are also shown in
Table~\ref{tab:farir}. These data include the recent estimates in the
far-IR from the Herschel-Atlas survey (Eales et al.~2009) and derived
from the same dataset used by Bourne et al., (2012) in generating
their Table~1. The galaxies from Bourne et al., in the redshift range
$0.01 < z < 0.12$, were k- and e-corrected to redshift zero. The k-
and e- corrections were derived from the stacked data and then applied
to each individual galaxy prior to stacking: k-corrections were
derived from the shape of the stacked SED in the redshift bin, while
e-corrections were based on the evolutionary fit to luminosities in
five redshift bins at $z<0.35$, given by $L(z) \propto
(1+z)^4$. Following the stacking of all optically-detected galaxies we
obtain fluxes of $(5.6 \pm 0.4) \times 10^{33}$ WMpc$^{-3} h_{100}$ at
250 $\mu$ m, $(2.1 \pm 0.1) \times 10^{33}$ WMpc$^{-3} h_{100}$ at 350
$\mu$ m, and $(6.1 \pm 0.4) \times 10^{32}$ WMpc$^{-3} h_{100}$ at 500
$\mu$ m. Although these values potentially miss a small amount of
faint emission the expectation is that this will be within the
errors. The errors quoted include errors from the stacking process and
a systematic error of 7\% due to the SPIRE flux calibration
uncertainty (see Pascale et al. 2012).

The match between our predicted far-IR CSED and the available data is
remarkably good, and corroborate our earlier conclusion (Driver et
al.~2008) that at low redshift ($z<0.1$) the dominant source of dust
heating is from attenuated starlight which is reradiated in the
far-IR.

\begin{figure}

\centerline{\psfig{file=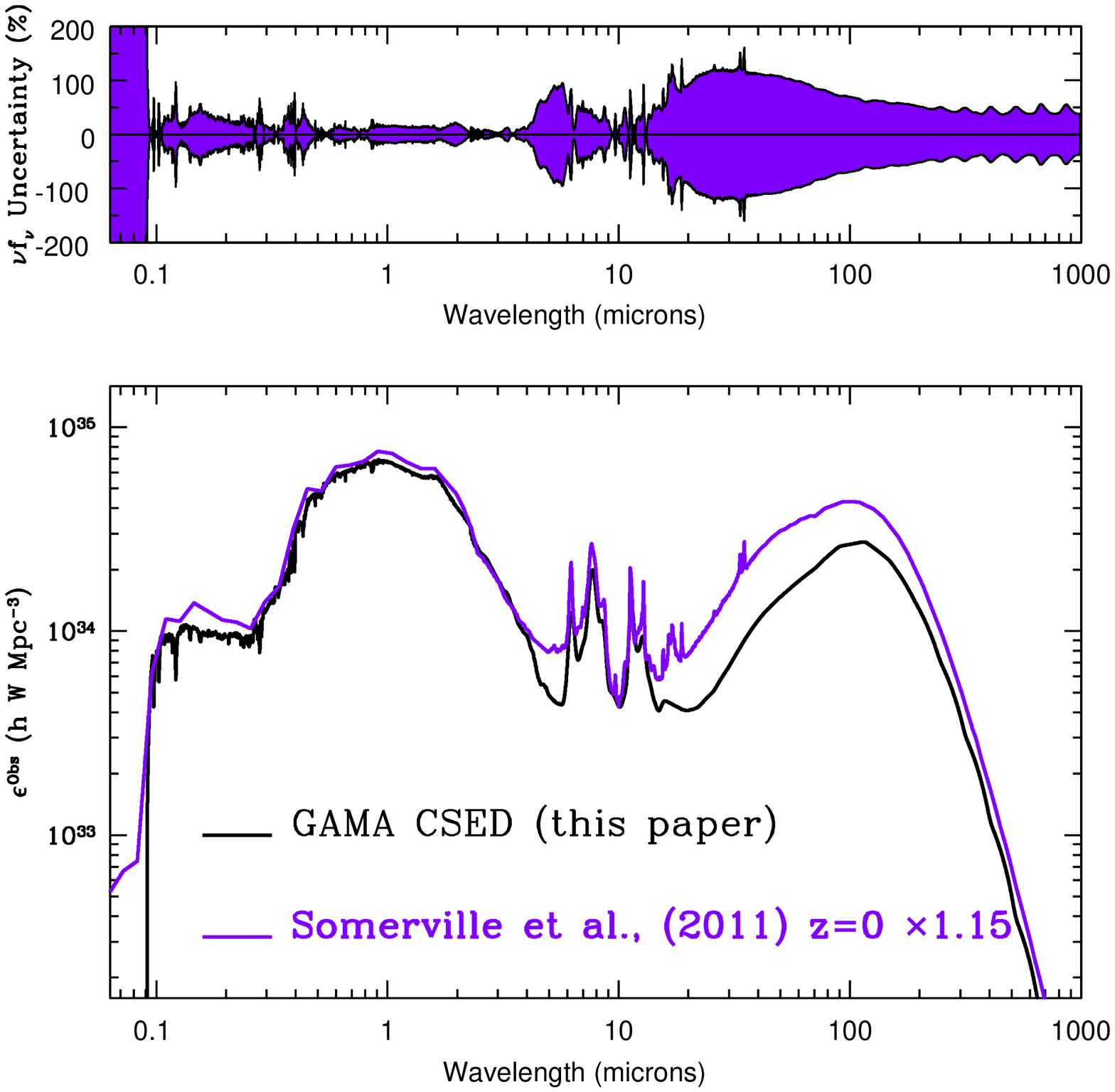,width=\columnwidth}}

\caption{\label{fig:energy2} (main panel) A direct comparison of the
  Somerville et al.~(2012) z=0 CSED and that derived in this paper.
  (upper panel) The ratio of the above curves indicating the wavelength
  regimes where there is large uncertainty in the models. The two
  curves have been normalised at 6175\AA ($r$-band).}
\end{figure}

Finally in Fig.~\ref{fig:energy2} we show a direct comparison of our
empirically constructed CSED with that modeled by Somerville et
al.~(2012). In general the two curves agree within a factor of 2 with
the largest discrepancies occurring in the mid and far-IR bands
indicating the regions of greatest uncertainty in our understanding of
the $z=0$ energy budget. Data from the WISE mission and Herschel-Atlas
(PACS) survey will be used in a follow-up paper to better constrain
the CSED over these wavelength ranges.

\begin{table*}
\begin{center}
\caption{The cosmic spectral energy distribution of the zero redshift
  Universe for various datasets and totals in units of h W
  Mpc$^{-3}$. The cosmic variance error is an additional $\pm 5$ per
  cent affecting all datapoints in a systematic manner. The errors
  shown for the dust corrected data include the uncertainty in the
  dust correction given in Col. 5 of Table~3. Note that Col. 4
  represents the luminosity density derived from the full sample not
  the summation of Cols. 2 \& 4. However Col. 6 represents the
  summation of Cols. 2 \& 6 and represents the final unattenuated
  CSED. \label{tab:csed}.}
\begin{tabular}{cccccc} \hline
Wavelength & Spheroids & Discs & Total & Discs (dust corrected) & Total (dust corrected) \\ 
\multicolumn{6}{c}{$\times 10^{34}$ W h Mpc$^{-3}$} \\ \hline
FUV (0.1535) & $(0.038 \pm 0.003) $ &  $(1.2 \pm 0.1) $ & $(1.2 \pm 0.1) $ & $(5 \pm 2) $ & $(5 \pm 2) $ \\ 
NUV (0.2301) & $(0.058 \pm 0.005) $ &  $(1.1 \pm 0.1) $ & $(1.2 \pm 0.1) $ & $(3 \pm 1) $ & $(3 \pm 1) $ \\ 
u (0.3557)   & $(0.36 \pm 0.02) $   &  $(1.74 \pm 0.08) $ & $(2.1 \pm 0.1) $ & $(3.8 \pm 0.7) $ & $(4.1 \pm 0.8) $ \\ 
g (0.4702)   & $(1.17 \pm 0.03) $   &  $(3.6 \pm 0.1) $   & $(4.7 \pm 0.1) $   & $(6.2 \pm 0.9) $ & $(7.4 \pm 0.9) $ \\ 
r (0.6175)   & $(1.76 \pm 0.05) $   &  $(4.5 \pm 0.1) $   & $(6.3 \pm 0.2) $   & $(7.5 \pm 1.1) $ & $(9.3 \pm 1.1) $ \\ 
i (0.7491)   & $(2.00 \pm 0.06) $   &  $(4.6 \pm 0.1) $   & $(6.6 \pm 0.2) $   & $(7.1 \pm 0.9) $ & $(9.0 \pm 1.0) $ \\ 
z (0.8946)   & $(2.11 \pm 0.06) $   &  $(4.6 \pm 0.1) $   & $(6.7 \pm 0.2) $   & $(6.6 \pm 0.7) $ & $(8.8 \pm 0.8) $ \\ 
Y (1.0305)   & $(1.99 \pm 0.09) $   &  $(4.1 \pm 0.2) $   & $(6.0 \pm 0.3) $   & $(5.7 \pm 0.7) $ & $(7.7 \pm 0.8) $ \\ 
J (1.2354)   & $(1.84 \pm 0.08) $   &  $(3.6 \pm 0.2) $   & $(5.4 \pm 0.3) $   & $(4.7 \pm 0.5) $ & $(6.5 \pm 0.6) $ \\ 
H (1.6458)   & $(1.81 \pm 0.08) $   &  $(3.5 \pm 0.2) $   & $(5.4 \pm 0.3) $   & $(4.3 \pm 0.4) $ & $(6.1 \pm 0.5) $ \\ 
K (2.1603)   & $(1.03 \pm 0.05) $   &  $(2.0 \pm 0.1) $ & $(3.0 \pm 0.2) $   & $(2.3 \pm 0.2) $ & $(3.3 \pm 0.2) $ \\  \hline
\end{tabular}
\end{center}
\end{table*}

\begin{table*}[h]
\caption{Summary of available far-IR CSED measurements at $z<0.1$
  which together with Table~\ref{tab:csed} constitute the empirical
  data shown on Fig.~\ref{fig:energy}. \label{tab:farir}}
\begin{tabular}{cccl} \hline
Wavelength & Dust emission & Facility & Reference \\
($\mu$m) & $\times 10^{34}$ W h Mpc$^{-3}$ & & \\ \hline
5.8 & $(1.2 \pm 0.1)$ & Spitzer & Babbedge et al.~(2006) \\
8 & $(2.6 \pm 0.2)$& Spitzer & Babbedge et al.~(2006) \\
8 & $(1.0 \pm 0.1)$& Spitzer & Huang et al.~(2007) \\
12 & $(0.69 \pm 0.07)$ & PCSz/IRAS/ISO & Takeuchi et al.~(2006) \\
25 & $(0.54 \pm 0.05)$ & PCSz/IRAS/ISO & Takeuchi et al.~(2006) \\
60 & $(1.6 \pm 0.2)$ & PCSz/IRAS/ISO & Takeuchi et al.~(2006) \\
100 & $(3.2 \pm 0.3)$ & PCSz/IRAS/ISO & Takeuchi et al.~(2006) \\
170 & $(2.0 \pm 0.2)$ & PCSz/IRAS/ISO & Takeuchi et al.~(2006) \\
250 & $(0.56 \pm 0.04$ & Herschel SPIRE & Bourne et al.~(2012) \\
350 & $(0.21 \pm 0.01$ & Herschel SPIRE & Bourne et al.~(2012) \\
500 & $(0.061 \pm 0.004$ & Herschel SPIRE & Bourne et al.~(2012) \\ 
850 & $(0.0036 \pm 0.0004)$ & PCSz/IRAS/ISO & Takeuchi et al.~(2006) \\ \hline
\end{tabular}
\end{table*}

\section{Conclusions}
Here we have used data from the GAMA survey (Driver et al.~2011) to
provide measurements of the galaxy luminosity function in 11 bands
($FUV,NUV,ugrizYJHK$) for a low-redshift ($z<0.1$) common imaging
region. This has enabled the construction of the cosmic spectral
energy distribution in unprecedented detail by producing a CSED
internally robust to intrinsic cosmic variance from 0.1 to 2.2$\mu$m
and absolute cosmic variance of $\pm 5$ per cent. The results
presented represent an order of magnitude improvement over
measurements compiled from the literature and offer the possibility to
provide significant new empirical constraints for numerical
models. The data essentially confirm our previous measurement however
five key results are highlighted.

~

\noindent
{\bf 1.} Colour is not a good separator of spheroid-dominated and
disc-dominated systems with significant contamination of the red
population by edge-on discs and the existence of blue spheroids. While
S\'ersic index appears to be a better discriminator there is still
clearly some ambiguity and no single definitive cut. We conclude
bulge-disc decomposition (e.g., Allen et al.~2007; Cameron et
al.~2009) is essential to robustly disentangle the two components.

~

\noindent
{\bf 2.} There appears to be a steep down-turn in the CSED at the
optical/near-IR boundary which appears real and prior to $Y$ band data
may have given rise to the appearance of an optical/near-IR
discontinuity. However that the $Y$ band data sits on a linear
extrapolation of the $z$ to $J$ band data this would imply that the
rapid decline across the optical/near-IR boundary is real. Current
state-of-the art models appear to struggle to follow this decline
which may imply the need for more detailed modeling of the stellar
evolution assumptions in this wavelength region.

~

\noindent
{\bf 3.} From our FUV data alone we derive a star-formation rate at a median
redshift of $z=0.078$ of $0.034\pm 0.003$(Random)$ \pm
0.009$(Systematic, Dust) $\pm 0.002$(Systematic, cosmic variance) h
M$_{\odot}$ yr$^{-1}$ Mpc$^{-3}$. 

~

\noindent
{\bf 4.} ($35 \pm 3$) per cent of the energy generated by stars in the
nearby Universe is attenuated by dust and reradiated in the
far-IR. This value is somewhat lower than our previous estimate ($45
\pm 5$ per cent) and although consistent within the errors the trend
is mainly due to the extension of our work into the NUV and FUV
bands.

~

\noindent
{\bf 5.} Using the UV to near-IR CSED and the azimuthally averaged photon
escape fractions we are now able to {\it predict} to a high degree of
accuracy the complete mid-IR and far-IR energy output in the nearby
Universe. The clear implication is that starlight is the dominant
source of dust heating in the nearby Universe. After correcting for
dust attenuation we find that the Universe is currently producing
energy at a rate of $(1.8 \pm 0.3) \times 10^{35}$ W h Mpc$^{-3}$.

~

In the coming years superior data from VST and VISTA along with more
extensive data in the mid-IR (WISE) and far-IR (Herschel PACS) which
should allow for further improvements of the CSED. Anyone wishing to
obtain either our pre- or post- attenuated CSEDs should contact
Simon.Driver@icrar.org.

GAMA is a joint European-Australasian project based around a
spectroscopic campaign using the Anglo-Australian Telescope. The GAMA
input catalogue is based on data taken from the Sloan Digital Sky
Survey and the UKIRT Infrared Deep Sky Survey. Complementary imaging
of the GAMA regions is being obtained by a number of independent
survey programs including GALEX MIS, VST KIDS, VISTA VIKING, WISE,
Herschel-ATLAS, GMRT and ASKAP providing UV to radio coverage. GAMA is
funded by the STFC (UK), the ARC (Australia), the AAO, and the
participating institutions. The GAMA website is
http://www.gama-survey.org/ .

Funding for the SDSS and SDSS-II has been provided by the Alfred
P. Sloan Foundation, the Participating Institutions, the National
Science Foundation, the U.S. Department of Energy, the National
Aeronautics and Space Administration, the Japanese Monbukagakusho, the
Max Planck Society, and the Higher Education Funding Council for
England. The SDSS Web Site is {\tt http://www.sdss.org/}.  The SDSS is
managed by the Astrophysical Research Consortium for the Participating
Institutions. The Participating Institutions are the American Museum
of Natural History, Astrophysical Institute Potsdam, University of
Basel, University of Cambridge, Case Western Reserve University,
University of Chicago, Drexel University, Fermilab, the Institute for
Advanced Study, the Japan Participation Group, Johns Hopkins
University, the Joint Institute for Nuclear Astrophysics, the Kavli
Institute for Particle Astrophysics and Cosmology, the Korean
Scientist Group, the Chinese Academy of Sciences (LAMOST), Los Alamos
National Laboratory, the Max-Planck-Institute for Astronomy (MPIA),
the Max-Planck-Institute for Astrophysics (MPA), New Mexico State
University, Ohio State University, University of Pittsburgh,
University of Portsmouth, Princeton University, the United States
Naval Observatory, and the University of Washington.

The UKIDSS project is defined in Lawrence et al.~(2007). UKIDSS uses
the UKIRT Wide Field Camera (WFCAM; Casali et al~2007). The
photometric system is described in Hewett et al.~(2006), and the
calibration is described in Hodgkin et al.~(2009). The pipeline
processing and science archive are described in Irwin et al.~(2008)
and Hambly et al.~(2008). We have used data from the 4th data release.

This research has made use of the NASA/IPAC Extragalactic Database
(NED), which is operated by the Jet Propulsion Laboratory, California
Institute of Technology, under contract with the National Aeronautics
and Space Administration.

\section*{References}

\reference Baldry I.K., Glazebrook K., 2003, ApJ, 593, 258

\reference Baldry I.K., et al., 2005, MNRAS, 358, 441

\reference Baldry, I.K., et al., 2010, MNRAS, 404, 86

\reference Baldry I.K., et al., 2012, MNRAS, 421, 621

\reference Bell E.F., McIntosh D.H., Katz N., Weinberg M.D., 2003, ApJS, 149, 289

\reference Bertin E., Arnouts S., 1996, A\&AS, 117, 393

\reference Blanton M.R., Roweis, S., 2007, AJ, 113, 734

\reference Blanton M., et al., 2003, ApJ, 592, 819

\reference Bourne N., et al., 2012, MNRAS, 421, 3027

\reference Budavari T., et al., 2005, ApJ, 619, 31

\reference Cameron E., Driver S.P., Graham A.W., Liske J., 2009, ApJ,
699, 105

\reference Cole S. et al., 2001, MNRAS, 326, 255

\reference Cook M., Lapi A., Granato G.L., 2009, MNRAS, 397, 534

\reference Cook M., Barausse E., Evoli C., Lapi A., Granato G.L., 2010, MNRAS, 402, 2113

\reference Cook M., Evoli C., Barausse E., Granato G.L., Lapi A., 2010, MNRAS, 402, 941

\reference Cross N.J.G., Driver S.P., 2002, MNRAS, 329, 579

\reference De Propris R., Liske J., Driver S.P., Allen P.D., Cross N.J.G., 2005, MNRAS, AJ, 130, 1516

\reference De Propris R., Conselice C., Liske J., Driver S.P., Patton D.R., Graham A.W., Allen P.D., 2007, ApJ, 666, 212
 
\reference De Propris R., et al., 2010, AJ, 139, 794

\reference Driver, S.P., Liske J., Cross N.J.G., De Propris R., Allen P.D., 2005, MNRAS, 360, 81

\reference Driver S.P., et al., 2006, MNRAS, 368, 414

\reference Driver S.P., Popescu C.C., Tuffs R.J., Liske J., Graham A.W., Allen P.D., De Propris, R, 2007, MNRAS, 379, 1022

\reference Driver S.P., Popescu C.C., Tuffs R.J., Graham A.W., Liske J., Baldry I., 2008, ApJ, 678, 101

\reference Driver S.P., et al., 2009, A\&G, 50, 12 

\reference Driver S.P., Robotham A.S.G., 2010, MNRAS, 407, 2131 (arXiv:1005.2538)

\reference Driver S.P., et al., 2011, MNRAS, 413, 971

\reference Driver S.P., et al., 2012, MNRAS, submitted

\reference Eke V.R., et al., 2004, MNRAS, 348, 866

\reference Fioc M., Rocca-Volmerange B., 1997, A\&A, 326, 950

\reference Fioc M., Rocca-Volmerange B., 1999, (arXiv:9912179) 

\reference Finke J.D., Razzaque S., Dermer C., 2010, ApJ, 712, 238

\reference Fukugita M., Hogan C.J., Peebles P.J.E., ApJ, 503, 518

\reference Gadotti D., 2009, MNRAS, 393, 1531

\reference Gallazi A., et al., 2005, MNRAS, 362, 41

\reference Gilmore R.C., Somerville R.S., Primack J.R., Dominquez A., 2012, MNRAS, 422, 3189

\reference Graham A.W., Driver S.P., Petrosian V., Conselice C.,
Bershady M.A., Crawford S.A., Tomotsugu G., 2005, AJ, 130, 1535

\reference Henriques B., Maraston C., Monaco P., Fontanot F., Menci N., De Lucia G., Tonini C., 2011, MNRAS, 415, 3571

\reference Hill, D., Driver S.P., Cameron E.C., Cross N.J.G., Liske J., 2010, MNRAS, 404, 1215

\reference Hill, D., et al., 2011, MNRAS, 412, 765

\reference Haung J,-S., Glazebrook K., Cowie L.L., Tinney C., 2003, ApJ, 584, 203

\reference Hopkins A.M., Beacom J.F., 2006, ApJ, 651, 142

\reference Irwin M., 2008, in ESO Astrophysics Symposium on {\it The
  2007 ESO Instrument Calibration Workshop}, 542 (Publ: Springer)

\reference Jones D.H., Peterson B.A., Colless M., Saunders W., 2006, MNRAS, 369, 25

\reference Kelvin L. et al., 2012, MNRAS, 421, 1007

\reference Kennicutt, R.C. Jr., 1998, ARA\&A, 36, 189

\reference Kochaneck C., Pahre M.A., Falco E.E., Huchra J.P., Mader J.,
Jarrett T.H., Chester T., Cutri R., Schneider S.E., 2001, ApJ, 560,
566

\reference Komatsu E., et al., 2011, ApJS, 192, 18

\reference Kroupa P, 2002, Science, 295, 82

\reference Lawrence A., et al., 2007, MNRAS, 379, 1599
 
\reference Liske J., Lemon D.J., Driver S.P., Cross N.J.G., Couch W.J., 2003, MNRAS, 344, 307

\reference Loveday J., et al., 2012, MNRAS, 420, 1239

\reference Maraston C., 2005, MNRAS, 362, 799

\reference Maraston C., 2011, invited review (arXiv: 1104.0022)

\reference M\"ollenhoff C., Popescu C.C., Tuffs R.J., 2006, A\&A, 456, 941

\reference Montero-Dorta A.D., Prada F., 2009, MNRAS, 399, 1106

\reference Nelan J.E., et al., 2005, ApJ, 632, 137 

\reference Norberg P., et al., 2002, MNRAS, 336, 907

\reference Oke J.B., Gunn J.E., 1983, ApJ, 266, 713

\reference Pascale, E., et al., 2011, MNRAS, 415, 911

\reference Peng, C., Ho Luis C., Impey C.D., Rix H-W., 2010, AJ, 139, 2097

\reference Popescu C.C., Tuffs R.J., 2002, MNRAS, 335, 41

\reference Popescu C.C., Tuffs R.J., Dopita M.A., Fischera J., Kylafis N.D., Madore B.F., 2011, A\&A, 527, 109

\reference Prescott M., Baldry I.K., James P.A., 2009, MNRAS, 397, 90

\reference Richards G., et al. 2006, AJ, 131, 2766

\reference Robotham A.S.G., et al., 2010, PASA, 27, 76

\reference Robotham A.S.G., Driver S.P., 2011, MNRAS, 413, 2570

\reference Rowlands K., et al., 2012, MNRAS, 419, 2545

\reference Schechter P., 1976, ApJ, 203, 297

\reference Schlegel D.J., Finkbeiner D.P., Davis M., 1998, ApJ, 500, 525

\reference Schmidt M., 1968, ApJ, 151, 393

\reference Smith A.J., Loveday J., Cross N.J.G., 2009, MNRAS, 397, 868

\reference Somerville R.S., Gilmore R.C., Primack J.R., Dominguez A., 2012, MNRAS, 423, 1992

\reference Tasca L.A.M., White S.D., 2011, A\&A, 530, 106 

\reference Wright E., 2001, ApJ, 556, 17

\reference Wyder T.K., et al., 2005, ApJ, 619, 15

\reference York D., et a., 2000, AJ, 120, 1579

\reference Zucca E. et al., 1997, A\&A, 326, 477

\label{lastpage}

\end{document}